# A survey of trends and motivations regarding Communication Service Providers' metro area network implementations

***Abstract** – Relevance of research on telecommunications networks is predicated upon the implementations which it explicitly claims or implicitly subsumes. This paper supports researchers through a survey of Communications Service Providers' current implementations within the metro area, and trends that are expected to shape the next-generation metro area network. The survey is composed of a quantitative component, complemented by a qualitative component carried out among field experts. Among the several findings, it has been found that service providers with large subscriber base sizes, are less agile in their response to technological change than those with smaller subscriber base sizes: thus, copper media are still an important component in the set of access network technologies. On the other hand, service providers with large subscriber base sizes are strongly committed to deploying distributed access architectures, notably using remote access nodes like remote OLT and remote MAC-PHY. This study also shows that the extent of remote node deployment for multi-access edge computing is about the same as remote node deployment for distributed access architectures, indicating that these two aspects of metro area networks are likely to be co-deployed.*

Etienne-Victor Depasquale [a,1], Mark Tinka [b], Saviour Zammit [a] and Franco Davoli [c]

[a] Department of Communications and Computer Engineering, Faculty of ICT, University of Malta, Msida MSD2080, Malta

[1] E-mail (corresponding author): edepa@ieee.org

[b] Head of Engineering, SEACOM, Level 1, Block A, Anslow Office Park, 8 Anslow Crescent, Anslow Lane, Bryanston, South Africa

[c] Department of Electrical, Electronic and Telecommunications Engineering and Naval Architecture, University of Genoa, Via All'Opera Pia, 11 A, 16145 Genoa, Italy

## 1. Introduction

Implementations of telecommunications networks span a multi-dimensional diversity of layers, technologies, topologies and architectures. The concreteness afforded by implementations is necessary to any research that requires the support of one or more scenarios to provide a context within which to state its aims. However, the task of drafting a scenario within this multi-dimensional space poses a daunting challenge. For example: should study span end-to-end communication? Which are the end systems at the end points of communication? Which protocols regulate communication? Of these, do any protocols dominate practice pertinent to the goal of the study? What is the current state of the perennial tussle between circuit- vs packet-switching, and connection-orientation vs connectionless-orientation? What level of abstraction retains those intermediate systems, or other artefacts, that are pertinent to the goal of the study? Which frameworks can be brought to bear to inform the researcher in the process of making these decisions? These decisions are an integral part of the study. At best, choices made constrain the relevance of research; at worst they render it misleading. In other words, the best-case scenario is a statement of research delimitations; such ***delimitations*** are a universal condition of research. The worst-case scenario is a presentation of research without acknowledgment of research *limitations*.

Guidance on these decisions comes from several sources, and one common method for harvesting guidance is through survey. However, a study of implemented telecommunications networks requires that the survey be construed more broadly than the ordinary confines of academic literature from conference proceedings and periodical journals. The compilation of a set of relevant *scenarios* depends on surveying that leads to understanding of current practice. In this extended sense, in this survey, data is collected through questionnaires (*quantitative survey*) and interviews (*qualitative survey*). When used jointly, the qualitative survey supports the analysis of the results obtained through the quantitative survey.

Notwithstanding the usefulness of these techniques, the multiple dimensions (layers, technologies, topologies and architectures) pose the challenge of how to design surveys of manageable length. A solution requires a different approach to the analysis of telecommunications networks: the *functional* approach, formalized in the *functional model* [1]. The functional model abstracts implementational[1] detail but systematically organizes the functions involved in telecommunications networks, within the context of the Global Information Infrastructure (GII). The principal categories of functions defined are Applications Functions, Middleware Functions and Baseware Functions [1]. The latter category is subdivided into Network Functions (NFs), Processing and Storage Functions (P&SFs) and Human-Computer Interfacing Functions (HCIF). NFs and P&SFs are found in network elements (NEs); these, therefore, are the objects of study where telecommunications networks are concerned. By focusing on such functions, a survey can digest the multi-dimensional space through the practical and highly useful technique of addressing dominant components, in any of the dimensions, within the functions. This follows because the surveyor can rely on rationalizing forces within Communications Service Providers (CSPs[2]) that reduces the multidimensional space to a few hotspots of interest that are, by virtue of their limited scope, far more tractable than the unconstrained space. In other words: by focusing on functions that CSPs are deploying, inferences can be made on the layers, technologies, topologies and architectures.

The scope of a survey of implementations can be refined further through consideration of another of the major concepts of the GII principles and framework architecture: the *implementational model.* The implementational[3] model emphasizes the physical organization of the telecommunications network. ITU-T Y.2011 aptly describes this as a concern with the way in which "functions are distributed and implemented in equipment" [2, p. 8]. The implementational model supports the formation of the scenario, which is defined in ITU-T Y.120 as "a combined graphical and textual representation of [a configuration of] … network technologies and user appliances that may be expected to be encountered in the context of the Global Information Infrastructure" [3].

---

[1] The adjective implementational has a formal meaning in the GII, and this will be introduced further on in this work, but its meaning here can be apprehended from the broader sense of the word "implementation".

[2] In this work, a CSP is defined as an organization that provides telecommunications services. Applications that use these services may also be offered by the organization as services themselves, but the defining attribute is the provision of telecommunications services.

[3] Within standardization documents regarding the ITU-T's Global Information Infrastructure (GII) (e.g., [1], [2]), both "implementation model" and "implementational model" are used.

A concern may arise about the relevance and usefulness of the implementational model, i.e., about the adoption of this type of model. An investigation of the Broadband Forum's (BBF) Technical Reports dispels any doubt. For example, TR-178 Issue 2 [4] ("Multi-service Broadband Network Architecture and Nodal Requirements") states the ***function modules*** comprising various types of access node (Ethernet-based-, MPLS-based-, and BNG (broadband network gateway)-embedded-access nodes) and describes various types of access node (AN) deployment *scenario*. Similarly, TR-470 Issue 2 [5] ("5G Wireless Wireline Convergence Architecture") is replete with scenarios, and makes extensive use of functional blocks, located within specific items of equipment (e.g., residential gateways (RGs) and ANs), to obtain a physical interpretation of 5G wireless-wireline convergence (WWC).

To date, a unified reference configuration of several wireline and wireless access technologies at the subscriber's end has been published in [6]. The scenarios unified under that reference configuration span the metro area network from the S reference point (RP) to the access node interface (ANI) RP (for optical distribution networks (ODNs) compliant with ITU-T G.984.1 [7], this is also equivalent to the S/R RP). The outstanding part of the pursuit of scenarios of CSPs' metro area telecommunications networks concerns the development of an implementational model of the complement to the access network in the metro area, extending from:

- the access node[4] at one end, to
- high-volume packet switches interfacing to long-haul links at the other end.

This "complement" may be succinctly referred to as the AN – metro-core span. The overarching goal of the remaining work, is, therefore, ***to provide a representative sample of implementational models of the AN – metro-core span***, thereby supplying directly usable implementational models, and enabling development of further models.

## 2. A three-axis framework for further development

The implementational model weaves ***topological components*** (see ITU-T G.805[8]), network functions, and processing and storage functions into a scenario suited to the purposes of the researcher requiring a context for a study's aims. Development of the implementational model must investigate the evolution of these two principal elements (topological components and functions) to accurately represent functional deployment within the metro area network.

---

[4] The access node is a key demarcation between the access portion and its complement in the metro area network. It is dealt with in detail in this work.

*2.1 Classical deployment of network, processing and storage functions*

The functions *classically* deployed in, and upstream of, the access node in the metro area, are the following.

- Transport aggregation (while going upstream; it is distribution in the downstream): Aggregation is a well-distributed function of specific network elements en route upstream from the access node. Within the scope of the metro area network, traffic thus aggregated terminates at one end on the user equipment (UE) and at the other end on a network gateway, e.g., an Internet BNG, a video BNG, an EPC SGW (serving gateway), and, in 5G, a user-plane function (UPF).
- Service authentication, authorization and accounting (AAA): this is localized but the equipment, perhaps in the form of a BNG, may be deployed in several places. The BNG may implement the AAA functions of a video service private to the CSP, or it might implement the AAA functions for the global Internet.
- Traffic classification (for service differentiation): this has traditionally been delegated to access node aggregation (L2) switches, with the support further upstream of provider edge (PE) packet switches (L3).

*2.2 Evolution of topological components*

Topological components are evolving under three major pressures.

a) ***The development of technologies that disrupt established topologies.*** A useful example regards XR optics. In XR optics, a frequency band is divided into (digital sub-) carriers; this facilitates the division of a single optical channel of large data rate capacity into several sub-channels. A passive splitter/combiner distributes the sub-channels to end-points for which lower data rates suffice. Thereby, the ***cost*** of distribution and aggregation is significantly reduced [9] (transceivers at the point of distribution/aggregation – the hub – are reduced from the number of end-points to one) through this point(one transceiver at the hub)-to-multipoint(one transceiver per end-point) arrangement.

b) ***The adoption of new technologies by communities of network operators (CSPs)***. The litmus test of technology is its adoption in the field. For example, while ATM used to be an aggregation architecture, its role has been supplanted by Ethernet (see, for example, the BBF's report on migration from ATM to Ethernet-based migration [10]). Each technology impacts either topological components, or location of deployment of network functions, or scope of deployment thereof, or any combination of the three.

c) ***Convergence of wireline and wireless networks, and convergence within wireline.*** Rationalization of diverse services onto a common transport infrastructure is a major motivational force for CSPs; the BBF's TR-470 (Issue 2) lists seven specific motives for convergence (of wireless and wireline) [5, Sec. 1.1]. Evidently, convergence re-writes all models; moreover, it is a major area of activity within ETSI, where Industry Specification Group (ISG) Fifth Generation Fixed Network (F5G) has, as of May 2023, published a network architecture with an explicit business requirement of "convergence and consolidation" of the various fixed networks [11, Sec. 4.2].

*2.3 The three-axis framework*

Therefore, three major axes for development of the implementational model can be perceived.

1. The implementation of classical functions is a mature field and current practice must be investigated and documented.
2. Due consideration of the three pressures' resultant thrust needs to be taken. These form trends that may be detectable; any detected trend must be documented and analyzed for its causes and effects.
3. The paradigmatic shifts of 5G and MEC introduce new participants (e.g., related to computing embedded in the MAN), new functions and new interconnection points. An observation that 5G and MEC should be added to the three pressures would be a good one. We have separated 5G and MEC from the set of three pressures because the latter predominantly influence topological components of the AN – metro core span, while 5G and MEC mostly introduce new functions to this span of the metro area.

A terse summary of the framework would thus be: depart from the status quo and selectively, through investigation of trends, include emerging functions and topologies. The three major axes form a framework that guides the formation of a method that can be brought to bear on the problem of an implementational model for the metro area of a telecommunications network. ***Quantitative*** and ***qualitative*** surveys align well with the three major axes. They support both investigation of implementation of classical network functions (1$^{st}$ major axis), as well as an investigation of CSPs' adoption of new technologies (2$^{nd}$ major axis) and functions (3$^{rd}$ major axis). In the interest of concision, a criticism of the use of questionnaires, a defence against the criticism and practical implications of the defence on the method, are delegated to an online supplement[5] to this work. The practical implications may be summarized as five guidelines supporting design of quantitative surveying of CSPs: (1) major market trends and shifts are reliably predicted by quantitative survey but narrower interests are less reliably predicted; (2) write unambiguous questions, using definitions where necessary; (3) ask qualified people: here, this means people in network operations and/or understand plans for operations;

---

[5] https://github.com/edepa/TrendsAndMotivationsInCSPMAN/blob/main/On%20the%20use%20of%20questionnaires%20as%20a%20reliable%20tool%20to%20gather%20data.pdf

(4) individually scrutinize responses for evident inconsistencies (and eliminate these responses); (5) a sample size of 80 – 100 responses provides statistics that match population parameters.

## 3. Surveying CSPs within the framework of the three major axes

### 3.1 The quantitative survey

#### The questions

The questionnaire underlying the quantitative survey has six optional, technical sections concerning metro area network architecture and one mandatory section concerning demographics. The organization of the technical sections is illustrated in Fig. 1. A brief description of the sections follows.

1. The first two (technical) sections concern size and rate of growth of access technologies by subscription, while the third concerns architecture of access technology implementation.
2. The fourth concerns the network services that CSPs are selecting to deploy the various stages of RAN traffic hauling (fronthaul, midhaul and backhaul), and the dissemination of use of disaggregated cell site gateways (DCSGs).
3. Aggregation (beyond the V RP) is dealt with in the fifth technical section. Concern here lies with trends and shifts in the optical network (layer $0^6$), the physical layer (layer 1) and the link layer (layer 2).
4. The sixth section is aimed at an investigation of the adoption of deep service edges (i.e., service edge closer to the subscriber than the local exchange/central office/HFC hub).

A copy of the questionnaire may be found [online](#)[7].

#### The sample

Two samples were collected. One of the samples was obtained through market research, conducted by SG Analytics [17]. SG Analytics has conducted several studies with CSPs. SGA's pool of respondents comprises all types of CSPs across the globe. The CSPs are differentiated based on the services they offer and the size of their company. For this study, global and regional CSPs from the pool like AT&T, T-Mobile and Verizon were targeted. Respondents were vetted by SGA before being included in the sample. SGA populates its pool of individuals from sources such as LinkedIn, Zoom info, conferences, and magazines. Data on pool individuals is processed using general criteria; individuals who pass these then proceed to project-specific criteria; this ensures that only qualified respondents participate in the study.

---

[6] An optical network takes the "layer 0" attribute when it comprises a set of components that is capable of selecting and switching a wavelength from source to destination. Such a network forms an additional layer as it provides services (channel capacity multiplication and lightpath variability) and has functions (selecting and switching wavelengths).

[7] https://forms.gle/QtoTkhzEk4Q1BLdVA

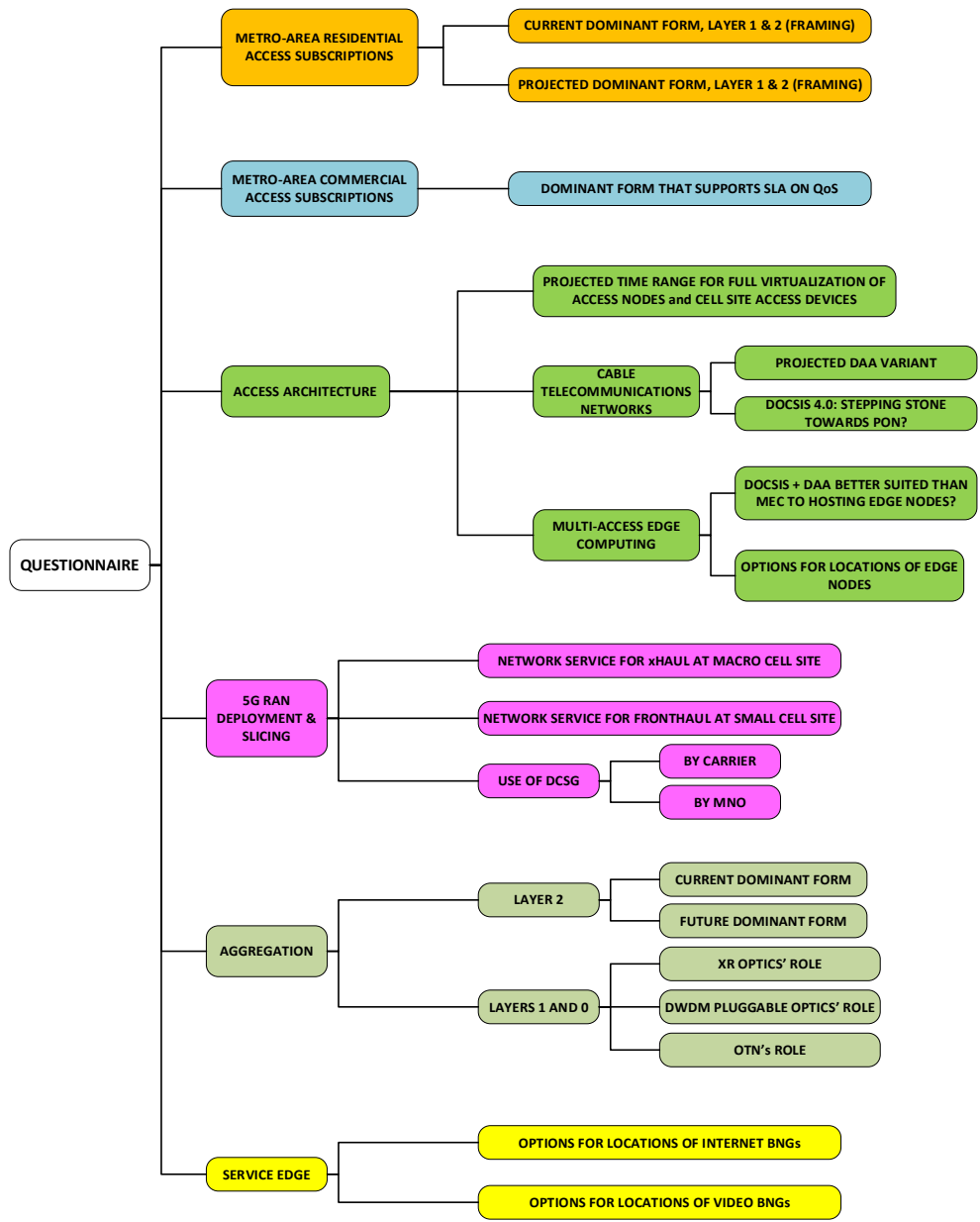

Fig. 1. Structure of the questionnaire used in the quantitative survey

- **General criteria**
    - ***Demographics*** Pool information is updated every three months, to capture the most recent status of the pool individual member's demographic details. The primary focus is on the individual's decision making in his/her current organization[8].
    - ***Behavioural patterns*** Every pool individual member is clubbed by SGA's scoring system, in order that his/her behavioural patterns on past projects may be assessed. Behavioural patterns are monitored by SGA's panel team on a regular basis.

---

[8] Demographic information includes company size, industry, job title and department.

- o *Anomalies in responses* SGA runs validation checks periodically on a pool individual member's historical data, to find anomalies in responses. Any pool individual member who fails these checks is removed from the pool.
- **Project-specific criteria**

    For a specific project (such as this one) a pool individual member passes from pool to sample through a double opt-in, thereby becoming a **respondent**. Respondents pass through:
    - o SGA's panel's initial screening parameters that are set for the project (covering the basic demographics that are required for the project), and
    - o our own screening criteria (the demographics section of the questionnaire).

Only respondents that clear both sets of screening parameters participate in the project. The general and the project-specific criteria are designed to select the best-fitting pool members to participate in the survey. Furthermore, individual responses were read and those which presented inconsistencies (e.g., fastest-growing access technology was not among the set of deployed access technologies) were removed from the sample.

The second sample was collected from operator groups'[9] mailing lists around the world. Respondents were not authenticated, but only one fake response was found and a respondent required considerable effort to answer the entire set of questions in a coherent and cogent manner. In this latter regard: during qualitative review, Philip Smith observed that "we are stuck with those who are willing to volunteer their time" [18]; the time estimated by SGA's personnel for a response to the complete set of questions is between 10 and 12 minutes [19].

### The quality of the questionnaire

The five recommendations are now brought to bear to assess the quality of the questionnaire.

1. **Major trends or minor movements?** The finest detail in the survey is found in the section asking about aggregation in the optical network. One question there asks about motives for migrating towards optical networks that use DWDM transceivers that are directly pluggable into router chassis faceplates. Otherwise, all questions regard differentiating between major technologies. For example, the aggregation section asks about current and future adoption of optical transport network (OTN, [20]) and about whether provider bridging (Q-in-Q) is preferred over seamless MPLS[10] transport; the service edge section asks about current locations where Internet BNGs are in use (and illustrates the locations using a graphic from the BBF's TR-178).

---

[9] In alphabetical order: AFNOG, APOPS, AUSNOG, DENOG, ENOG, FRNOG, GORE, IDNOG, INNOG, ITNOG, JANOG, LACNOG, NANOG, SAFNOG, SANOG, SWINOG, UKNOF

[10] Seamless MPLS is the evolutionary step wherein all aggregation within the CSP's network, from access node to access node, or from access node to E-NNI, takes place over MPLS label-switched paths.

2. **Write unambiguous questions, using definitions where necessary.** Apart from support during crafting, the qualitative survey (described later) asked interviewees specifically about whether they had detected ambiguities or lack of clarity. Three issues emerged.

   a) The term "routed optical network" is proprietary to Cisco. A discussion was opened in this regard on NANOG[11] [21]. Eduard Vasylenko[12] correctly identified this as term as proprietary; Ovidiu-Mădălin Roşeţ, a CCIE (Cisco Certified Internetwork Expert) opined that the term was not likely associated with Cisco's use thereof [22, N. @35:23]. These observations are mitigated by one from an interview [23, N. @8:36] with another CCIE (Haider Khalid) in the qualitative review phase, that "Cisco is actually the go-to vendor for everybody" (matching my personal experience). The consequences of this ambiguity are handled in the interpretation of the results.

   b) "Segmented (as opposed to seamless) MPLS transport" might not be universally understood; a doubt emerged in [22, N. @4:41]). However, the segmented pseudowire is well-defined [24], and use of the term in contraposition to seamless MPLS should have sufficed to distinguish the meaning intended. Moreover, in "draft-ietf-mpls-seamless-mpls-07" [25, Sec. 2.1], a clear contraposition is made between the status quo ante and the proposal for seamless MPLS.

   c) XR optics (see reference made earlier) is sufficiently novel to evade correct identification. During a review with Jon Baldry[13] in September 2022, Infinera's "Director Metro Marketing", doubt was cast on the claim that two respondents from the NOG sample had already deployed the technology, as he was unaware of any commercial deployments at the time of the survey. It is possible that the respondents were referring to XR optics field trials, rather than commercial deployments, as many of these trials had not yet been committed to deployment stage in the 2021-2022 timeframe. A clarifying reference was added to the question, but data collection was by then roughly at two-thirds the final sample's size.

From a more general perspective: two reviewers were explicitly invited to comment about limitations and ambiguities in the questionnaire's questions [26], [27]. They explicitly stated that they did not see any ambiguities. Both suggested some extension of scope of the questionnaire, but these were concerned with the control plane rather than the data plane.

---

[11] In all cases, "NOG" represents "network operators' group".
[12] https://www.linkedin.com/in/eduard-vasylenko-b723ab1
[13] https://www.linkedin.com/in/jon-baldry-5605b/

3. **Ask qualified respondents**. The problem of qualifying respondents was dealt with through two approaches.

   a) Relevant organizational roles were specified (see the demographics section of the questionnaire in the online resources[14]).

   b) A vetted database of respondents was used in the first sample.

4. **Scrutinize individual responses and eliminate those with inconsistencies**. Every response in both samples was subjected to scrutiny. This process is documented for the first sample in [16].

5. **Sample size of 80 – 100 responses**. The first sample was restricted to 50 responses because of budgetary constraints, with 30 different companies represented. The second sample consists of 79 responses; this should still suffice.

*3.2 The qualitative survey*

The objectives may be broadly classified under two headings.

1. Discuss the graphical summaries of the quantitative survey's results.
2. Assess the objective clarity of the questions in the questionnaire.

Qualitative survey was carried out by means of some or all of the following:

1. Face-to-face interviews, which were recorded;
2. e-mails, and
3. a written assessment.

Face-to face interviews

1. Both the objective classes were tackled during the interviews.
2. Three participants were recruited.

   a. Two were recruited by SGA through a process that included filtering by a set of screening questions which was written (see the online resources[15]) with the support of anonymous_2 (the network engineer).

   b. The latter's suggestion (in [12]) to recruit knowledgeable people with direct participation in the field of network operations – colloquially, we would say that their hands are dirty – was supported by Mark Tinka. Our own experience gives credence to this opinion; to strengthen it further, we qualified it (in the screening questions) by the criterion that prospective interviewees should be experienced.

---

[14]https://github.com/edepa/TrendsAndMotivationsInCSPMAN/blob/main/Questionnaire%20-%20quantitative%20survey.pdf
[15] https://github.com/edepa/TrendsAndMotivationsInCSPMAN/blob/main/Screeners%20for%20qualitative%20survey.pdf

3. All interviews were recorded [22], [23][16].

E-mails

1. Both the objective classes were tackled in the course of the exchanges in the mail threads.
2. Four highly-experienced and highly-qualified participants shared their interpretations:
    a. [Mark Tinka](https://www.linkedin.com/in/mark-tinka-5b03055/)[17];
    b. [Philip Smith](https://www.linkedin.com/in/philip-smith-154502/)[18];
    c. [Daniel King](https://www.linkedin.com/in/danielking/)[19];
    d. reviewer anonymous_2.
3. All interactions are documented and available [12], [18], [27], [28].

Written assessment

While interviews are a useful medium, they are limited by the time which all parties involved can endure in discussion and remain at sufficient ease to interpret graphical summaries of data. One interviewee accepted the request to follow up with a written assessment.

1. Both the objective classes were tackled during the written assessment.
2. Haider Khalid, a highly-qualified network engineer participated.
3. The assessment is documented and available [26].

*3.3 The data*

Both quantitative and qualitative surveys harvested data, and these data are presented in a Mendeley dataset [29]. The quantitative survey's data is presented in the form of two Excel spreadsheets, one of which gathers NOG respondents' answers, while the other gathers answers from SGA's respondents. The qualitative survey's data take two primary forms: one is a textual form, consisting of discussion threads that developed over the course of (sometimes lengthy) e-mail exchanges and the other is that of a video recording (of interviews). As regards the qualitative survey's data, permission for publication was sought, and, in most cases, granted. Some anonymization had to be effected to observe the instructions of those contributors who wished to remain anonymous or wished to preserve the privacy of parts of their contributions.

## 4. Quantitative survey results

This section presents a graphical summary for each question that concerns the AN – metro-core span, and for each question that concerns the service edge and 5G functions. Since MEC nodes (these form part of the deployment range for the service edge) and 5G functions may be deployed within the access network, questions that regard them (the MEC nodes and 5G functions) may pertain to the access network. The graphical summaries consist of histograms, bar charts, clustered bar charts and pie charts. Brief commentary accompanies the charts, to draw attention to noteworthy characteristics. Analysis is deferred to section 8.5, where, aided by reviewers' comments, decisions on which scenarios to model are taken (sub-section 8.5.1).

---

[16] We have not been given permission to publish the recording of anonymous_1's interview.
[17] https://www.linkedin.com/in/mark-tinka-5b03055/
[18] https://www.linkedin.com/in/philip-smith-154502/
[19] https://www.linkedin.com/in/danielking/

*4.1 An overview of both NOG and SGA samples*

This sub-section first presents a distribution of customer base size, and essential statistics thereof, for both samples. For brevity's sake, the sample collected from the operator groups will be referred to as the NOG sample, while the sample collected from SGA's database will be referred to as the SGA sample. Figures will be presented with the NOG chart on the left and the SGA chart on the right. It then proceeds to extract a correlation from the adoption of technologies in the access network (this segment is both CapEx and OpEx intensive); this correlation is important because it suggests an interpretive bias for the rest of the results. The sub-section concludes with a summary of the results, in Table II .

Essential statistics on subscriber base size

The results of the demographics questions are useful to understand respondents' organizations better. For every respondent, a conservative estimate of size of subscriber base was obtained as follows. Consider Fig. 2. Suppose that a respondent selects the radio button corresponding to a choice, say, in the range one thousand – one hundred thousand subscribers, in one – ten metro areas. Before aggregation with the other answers for the other geographical areas, this is converted (for the region of North America) to one thousand subscribers in one metro area – thus, a conservative estimate. This is then added to the conservative estimate for each other geographical area, to obtain a conservative estimate of the respondent's global subscriber base size. By processing each respondent's input in this way, **a set of conservative estimates** is obtained. The set of such conservative estimates provides a sample, and this sample was processed to extract the mean, median and mode statistics (Table I ) and the distribution of size of subscriber base (Fig. 1 shows two histograms; NOG sample on left, SGA sample on right).

Table I   ESSENTIAL STATISTICS ON SUBSCRIBER BASE

|  | **NOG sample (thousands of subs.)** | **SGA sample (thousands of subs.)** |
|---|---|---|
| Mean | 5510 | 42887 |
| Median | 100 | 30687 |
| Mode | 1 (28 instances) | 101 (2 instances) |

Page 12 of 70

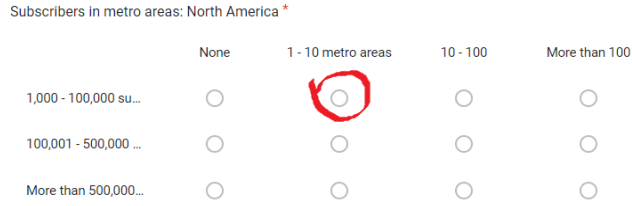

**Fig. 2.** Choice of 1,000 – 100,000 subscribers in 1 – 10 metro areas is reduced to a conservative value of 1,000 subscribers

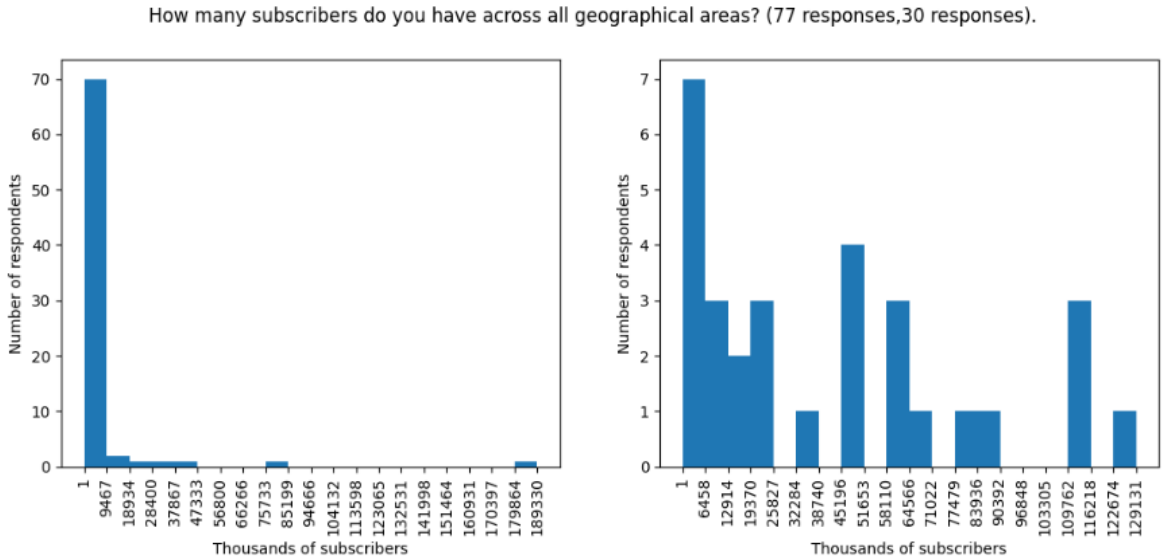

**Fig. 1.** Distribution of a conservative estimate of size of subscriber base

A correlation between subscriber base size and access network technology

The conservative approach proved very useful in explanation of differences between the responses from the two samples. The sizes reported by NOG respondents are heavily skewed towards the low end of the range [1,000 , 189,330,000]. The sizes reported by SGA respondents are less heavily skewed towards the low end of the range [1,000, 129,131,000]. This indicates that while SGA respondents do indeed (as claimed by SGA) include regional operators, the NOG respondents have smaller subscriber base size and more likely to be later entrants in the population of CSPs. This observation is reinforced by the questions regarding access technologies; while these are clearly not within the scope of the AN – metro-core span, the access network is the largest sink of capital and operational expenditure. Fig. 2 shows the adoption of access technologies by respondent. Inspection of Fig. 2 reveals that the NOG respondents – the smaller CSPs – claim that GPON is both the most adopted access technology (blue bars), as well as the one that is growing at the fastest rate (orange bars). On the other hand, the larger operators – captured in SGA's sample – identify ADSL2+ as most adopted and fastest growing access technology.

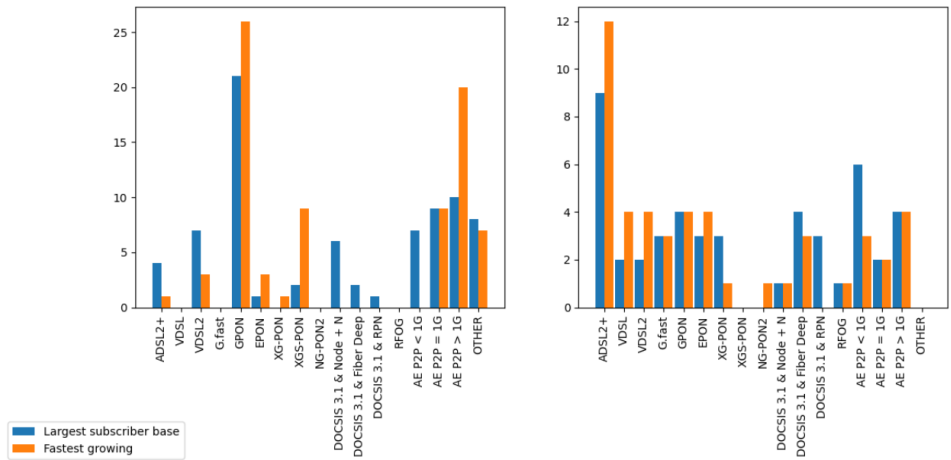

Fig. 2. Access technology adoption, unweighted

This result merits a deeper look before the full analysis is dealt with, as it serves as part of the interpretive lens through which the survey's data is to be read. Consider Fig. 3[20], which shows a *weighted* version of the responses, i.e., each response is multiplied by a factor numerically equal to the conservative estimate of the subscriber base's size. Clearly, not all a respondent's subscribers are on the same access technology; at the same time, if a respondent indicates that an access technology is the respondent's most adopted access technology, then further light might be shed. In the weighted set, ADSL 2+ is the largest and fastest growing in both samples. This indicates that the more established CSPs (including incumbents) are still largely exploiting their investments in copper media. This conclusion was reached following discussion with SGA and an expert reviewer [28]. Therefore, further interpretation of the data can be supported by the understanding that respondents in the NOG sample are likely to be less bound by legacy than those in the SGA sample. This understanding was affirmed in discussion with another expert reviewer [18].

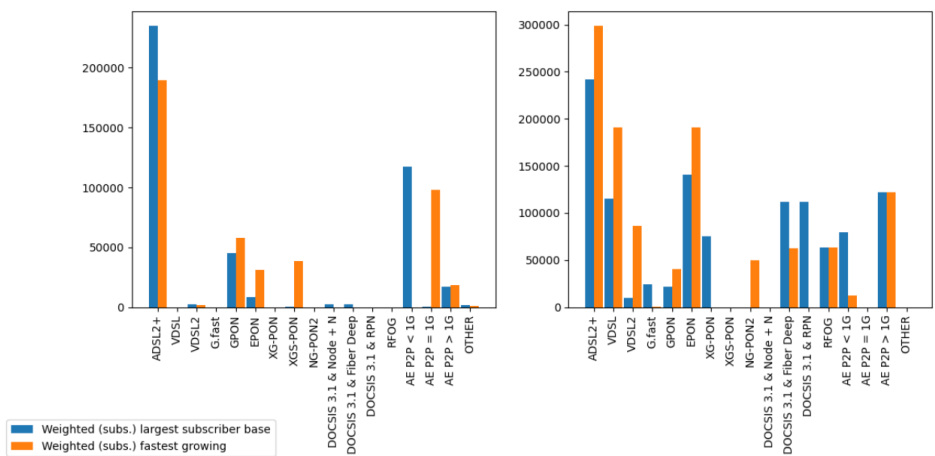

Fig. 3. Access technology adoption, weighted by size of subscriber base

---

[20] The figure only shows 28 respondents in SGA's sample; these correspond to the 28 different CSPs in the sample.

### A quick-and-dirty differentiator: Tier 1 vs "regional" vs "incumbent"

The role played by "Tier 1" CSPs is not identical to that played by the incumbents. Tier 1 CSPs provide transit to other CSPs and to service providers (e.g., video) who operate OTT, as well as to Cloud data centres. Incumbents, as stated earlier, are most likely to be CSPs who have invested heavily in structural and infrastructural works, possibly being the privatised descendant of erstwhile state-owned enterprises for national telecommunications. Incumbents may be restricted to regional (perhaps even national) operation. On the other hand, the Tier 1 CSPs, while quite possibly descended from an operator of regional/national scope, have a broader geographical (possibly global) scope. Examples of regional/national operators who have evolved to include the Tier 1 role, are Telefónica and Deutsche Telekom, but there are newer entrants like CSP_V[21] who do not share the same origin story. The latter do not have a significant residential service (this emerged from a direct interview with CSP_V senior technical personnel), whereas the former two do. Moreover, subscriber base size is not equivalent to revenue size. While residential subscriptions are well-known to be low-margin, the same cannot be said about enterprise services.

### 4.2 Access Architecture

What is the time range within which you plan to virtualize all your access nodes (vOLT vs OLT, vCMTS vs CMTS) and/or cell-site access devices (disaggregated cell site gateway (DCSG) vs cell-site router (CSR))?

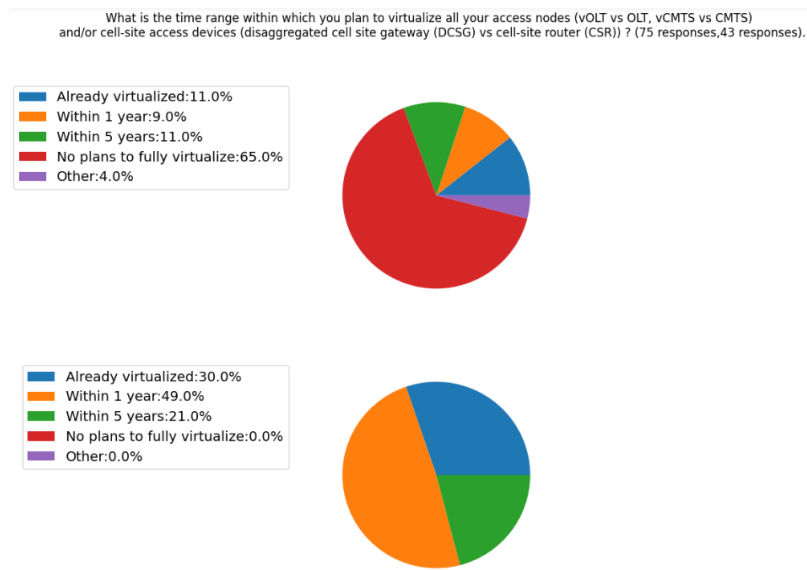

**Fig. 4.** Time span within which all access nodes and cell-site access devices will be fully virtualized (numbers show percentage of sample size)

Most NOG respondents do not plan to fully virtualize, while all SGA respondents plan to fully virtualize. This may reflect the relationship between the size of a CSP's employee cohort, and the cohort's skill set specialization. Larger CSPs are likely to employ more people and are likely to support greater specialization.

---

[21] Permission to reveal the name was not given. CSP_V is an operator of a global network, with a high density of points of presence (PoP) in European datacentres, and others in North America and the Far East.

## Tabulation of results of quantitative survey of CSPs

Table II SUMMARY OF RESULTS

**ACCESS ARCHITECTURE**

**Access node virtualization**

| GROUP | Done | <=1 y[22] | <=5 y | No plans to fully virtualize | Other | Responses |
|---|---|---|---|---|---|---|
| NOG | 11 | 9 | 11 | 65 | 4 | 75 |
| SGA | 30 | 49 | 21 | 0 | 0 | 43 |

**DAA option**

| GROUP | R-OLT | R-OLT conditional | RMN | RMN conditional | RPN | RPN conditional | Undecided, but within 5 years | No DAA | Other | Responses |
|---|---|---|---|---|---|---|---|---|---|---|
| NOG | 10 | 17 | 1 | 7 | 10 | 6 | 15 | 44 | 4 | 72 |
| SGA | 21 | 38 | 38 | 23 | 17 | 23 | 13 | 2 | 0 | 47 |

**DAA for majority HHP?**

| GROUP | <=2 y | <=5 y | Greenfield only | No | Other | Responses |
|---|---|---|---|---|---|---|
| NOG | 7 | 13 | 12 | 65 | 3 | 68 |
| SGA | 53 | 36 | 9 | 2 | 0 | 47 |

**Option 0 MEC node**

| GROUP | In progress | <=1 y | <=5 y | No plans to deploy | Other | Responses |
|---|---|---|---|---|---|---|
| NOG | 16 | 11 | 11 | 61 | 2 | 57 |
| SGA | 24 | 56 | 20 | 0 | 0 | 50 |

**Services for fronthaul at macro cell sites**

| GROUP | MEF service | MPLS service | PON ONU | Wavelength | Dark fibre | Wireless | Other | Responses |
|---|---|---|---|---|---|---|---|---|
| NOG | 14 | 32 | 5 | 12 | 32 | 2 | 4 | 56 |
| SGA | 24 | 35 | 16 | 8 | 3 | 14 | 0 | 37 |

**Services for midhaul at macro cell sites**

| GROUP | MEF service | MPLS service | PON ONU | Wavelength | Dark fibre | Wireless | Other | Responses |
|---|---|---|---|---|---|---|---|---|
| NOG | 11 | 39 | 2 | 13 | 27 | 2 | 7 | 56 |
| SGA | 22 | 24 | 5 | 19 | 8 | 22 | 0 | 37 |

**Services for fronthaul at small cell sites/fixed wireless access**

| GROUP | MEF service | MPLS service | PON ONU | Wavelength | Dark fibre | Other | Responses |
|---|---|---|---|---|---|---|---|

---

[22] This regards operators who plan to implement a feature within one year. Similarly, "<=5y" regards operators who plan to implement a feature within 5 years.

|  | NOG | 15 | 36 | 7 | 12 | 27 | 3 | 59 |
|---|---|---|---|---|---|---|---|---|
|  | SGA | 49 | 32 | 8 | 5 | 5 | 0 | 37 |

## AGGREGATION ARCHITECTURE

**Current dominant form of L2 or greater aggregation from access node (V RP) to service edge**

| GROUP | PB, w/o MPLS | Seamless MPLS | Segmented (not seamless MPLS) | PB near access, then MPLS | Other | Responses |
|---|---|---|---|---|---|---|
| NOG | 19 | 43 | 11 | 18 | 9 | 74 |
| SGA | 49 | 28 | 15 | 9 | 0 | 47 |

**Preferred form of L2 or greater aggregation from access node (V RP) to service edge**

| GROUP | PB, w/o MPLS | Seamless MPLS | Segmented (not seamless MPLS) | PB near access, then MPLS | Other | Responses |
|---|---|---|---|---|---|---|
| NOG | 14 | 51 | 8 | 17 | 10 | 71 |
| SGA | 43 | 30 | 19 | 9 | 0 | 47 |

**Do you plan to deploy XR optics in your metro-aggregation network?**

| GROUP | Already deployed | By end 2022 | By end 2023 | By end 2025 | Currently investigating | No plans | Responses |
|---|---|---|---|---|---|---|---|
| NOG | 4 | 1 | 7 | 5 | 27 | 55 | 74 |
| SGA | 0 | 26 | 21 | 6 | 40 | 6 | 47 |

**Existing OTN aggregation wil stay in my network but I won't choose OTN for any expansion of my aggregation network.**

| GROUP | Fully disagree | Somewhat disagree | Somewhat agree | Fully agree | Other | Responses |
|---|---|---|---|---|---|---|
| NOG | 9 | 33 | 46 | 12 | 0 | 57 |
| SGA | 4 | 2 | 70 | 23 | 0 | 47 |

**If you agree that you won't include OTN, why not?**

| GROUP | Cost | Granularity of b/w | Inability to meet 5G URLLC | Other | Responses |
|---|---|---|---|---|---|
| NOG | 52 | 19 | 21 | 8 | 36 |
| SGA | 48 | 29 | 22 | 0 | 44 |

**Packet-based networks will fully displace OTN from MANs, except in DCI.**

| GROUP | Fully disagree | Somewhat disagree | Somewhat agree | Fully agree | Other | Responses |
|---|---|---|---|---|---|---|
| NOG | 13 | 23 | 52 | 12 | 0 | 60 |
| SGA | 0 | 9 | 72 | 17 | 2 | 47 |

**Current dominant form of technology stack in metro aggregation**

| GROUP | DWDM+SDH+E+ IP/MPLS | DWDM+ROADM+ OTN+E+IP/MPLS | DWDM+ROADM+ E+IP/MPLS | DWDM+ROADM+ E+IP | Routed optical nets. over E without ROADM | Other | Responses |
|---|---|---|---|---|---|---|---|
| NOG | 10 | 13 | 14 | 6 | 45 | 12 | 69 |
| SGA | 23 | 23 | 28 | 13 | 13 | 0 | 47 |

**Greenfield form of technology stack in metro aggregation**

| GROUP | DWDM+ROADM+ OTN+E+IP/MPLS | DWDM+ROADM+ E+IP/MPLS | DWDM+ROADM+ E+IP | Routed optical nets. over E without ROADM | Other | Responses |
|---|---|---|---|---|---|---|
| NOG | 12 | 33 | 9 | 43 | 3 | 67 |
| SGA | 36 | 26 | 26 | 13 | 0 | 47 |

**Greenfield form of technology stack in metro core**

| GROUP | DWDM+ROADM+ OTN+E+IP | | DWDM+ROADM+ E+IP | Routed optical nets. over E without ROADM | | Responses |
|---|---|---|---|---|---|---|
| NOG | 10 | | 37 | 49 | 3 | 67 |
| SGA | 36 | | 40 | 23 | 0 | 47 |

**SERVICE EDGE**
**Service edge location for Internet BNG**

| GROUP | Option 0 | Option 1 | Option 2 | Option 3 | At A10 | Other | Responses |
|---|---|---|---|---|---|---|---|
| NOG | 19 | 33 | 46 | 27 | 19 | 0 | 32 |
| SGA | 4 | 36 | 44 | 34 | 8 | 0 | 50 |

**Service edge location for Video BNG**

| GROUP | Option 0 | Option 1 | Option 2 | Option 3 | At A10 | Other | Responses |
|---|---|---|---|---|---|---|---|
| NOG | 3 | 31 | 41 | 25 | 16 | 0 | 32 |
| SGA | 10 | 38 | 36 | 30 | 10 | 0 | 50 |

**Support for eMBB is improved by adding video BNGs closer to the end user**

| GROUP | Fully disagree | Somewhat disagree | Somewhat agree | Fully agree | Other | Responses |
|---|---|---|---|---|---|---|
| NOG | 2 | 22 | 50 | 20 | 5 | 40 |
| SGA | 0 | 4 | 48 | 48 | 0 | 50 |

**Energy efficiency is improved by adding video BNGs closer to the end user**

| GROUP | Fully disagree | Somewhat disagree | Somewhat agree | Fully agree | Other | Responses |
|---|---|---|---|---|---|---|
| NOG | 13 | 21 | 47 | 15 | 4 | 47 |
| SGA | 0 | 4 | 56 | 40 | 0 | 50 |

**Is Carrier Ethernet most adopted service on UNIs subject to QoS SLA?**

| GROUP | Yes | No | Responses |
|---|---|---|---|
| NOG | 67 | 33 | 78 |
| SGA | 41 | 2 | 43 |

Which distributed access architecture (DAA) option(s) are you planning for new deployments and replacement deployments?

Fig. 5 shows a bar chart of the NOG sample on top and one for the SGA sample below it. A pattern emerges similar to that which emerges from **Error! Reference source not found.**. Smaller operators are far less likely to commit to DAA than the incumbents. No other characteristics is readily perceived from the data. Technology choices are fairly equally distributed among all the candidate architectures.

Do you plan DAA to serve the majority of your households passed (HHPs, when compared with centralized access forms such as centralized OLT and integrated CCAP)?

Fig. 6 concurs with Fig. 5, in the sense that while 43% of the NOG sample do not plan to deploy any DAA option, even more (65%) do not plan to make it the majority access architecture. On the other hand, the overwhelming majority of incumbents (89%) plan to do so within 5 years.

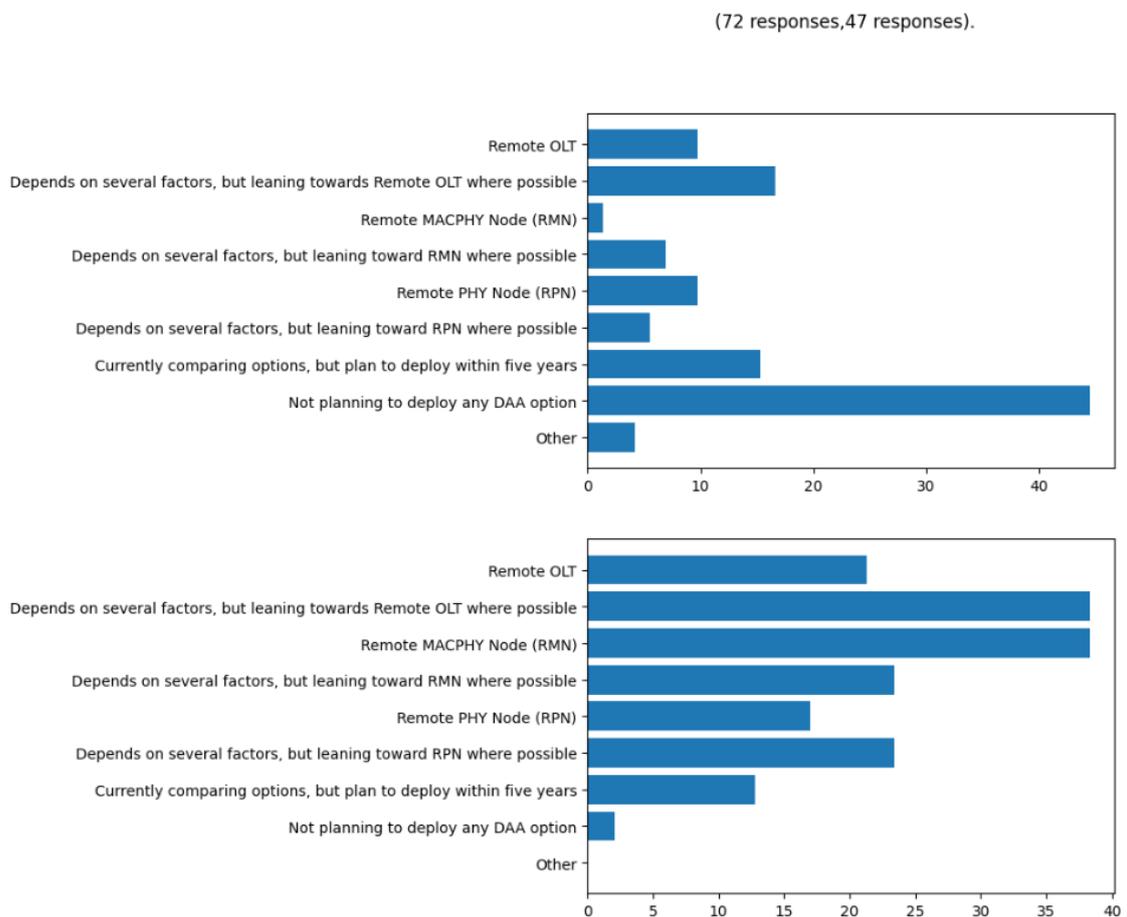

**Fig. 5.   Adoption of distributed access architecture technology (numbers show percentage of sample size, NOG on top)**

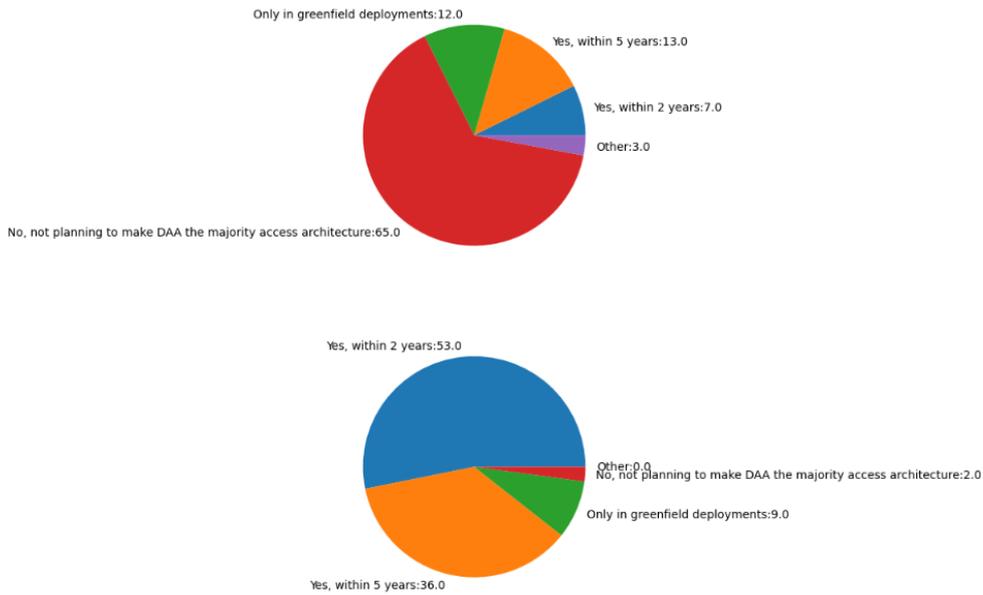

**Fig. 6.** Will DAA be your majority access architecture in HHPs? (numbers show percentage of sample size, NOG on top)

Do you plan to deploy remote access nodes (Option 0) to enable MEC services?

Option 0 (Fig. 7, [4, Fig. 2]) supports the greatest 5G and MEC functional range, but widespread provision is very capital-intensive (due to numbers). Results in the NOG sample (Fig. 8) are consistent with those shown in Fig. 6; similarly, SGA respondents seem intent on exploiting DAA real estate to deploy customer-proximal computing.

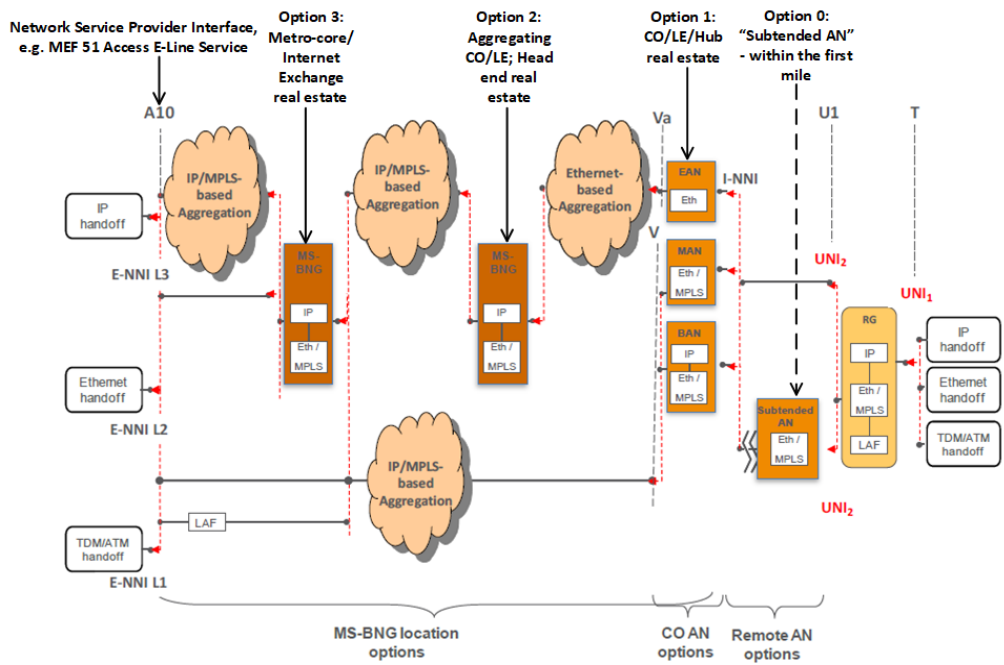

**Fig. 7.** General TR-178 architectural scheme, encompassing its targeted deployment scenarios [4, Fig. 2]

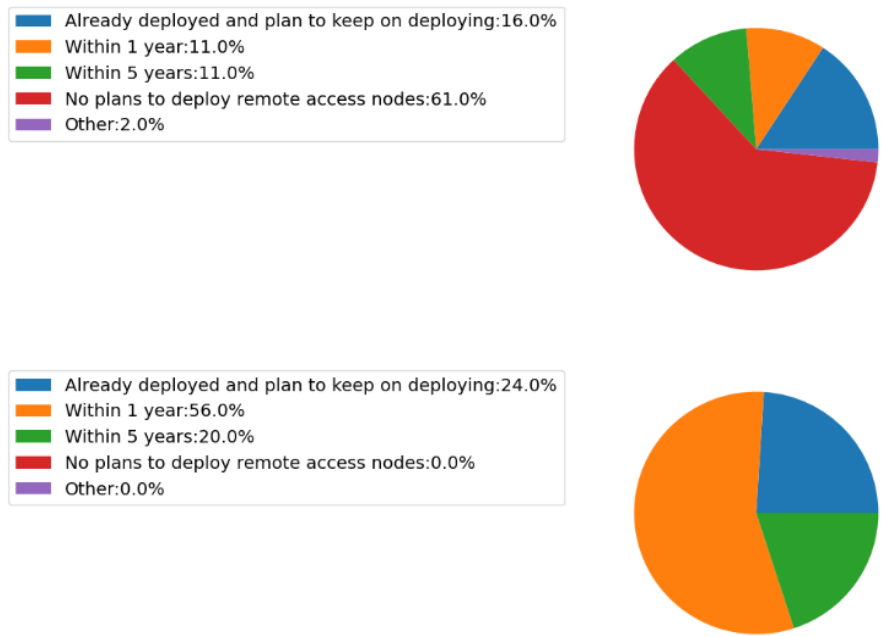

**Fig. 8.** Deployment of remote access nodes close to the customer (numbers show percentage of sample size, NOG on top)

For fronthaul / midhaul at macro cell sites, what type of network service have you deployed/purchased most commonly?

MPLS service dominates both samples for both fronthaul and midhaul, but dark fibre is a surprisingly strong contender in the NOG sample (Fig. 9). For the incumbents (SGA sample), MEF services are the second most adopted front- and mid-haul technology, but wireless mid-haul is equally popular in this sample. This is to be expected for incumbents, who would still be collecting return on their investment in microwave backhaul systems; such wireless front- and mid-haul is also well-known for use in rural centres distant from dense urban areas. The popularity of MEF service seems to emerge as a continuation of this class of technology's popularity as backhaul for 4G networks [30]. A MEF service is an Ethernet Virtual Connection (EVC), and it can be a point-to-point, point-to-multipoint or rooted-multipoint association between two (or more) MEF UNIs (this UNI is aligned with the T RP).

Note that the titular "network service" is a good example of CSPs' support of other CSPs (commonly referred to as mobile network operators, or MNOs) or of CSPs' own radio access network services. Moreover, while EVC and MPLS services are competitive, they also may be complementary; for example, an EVC may use MPLS for its transport (transport is recursive). This theme will be dealt with in the analysis.

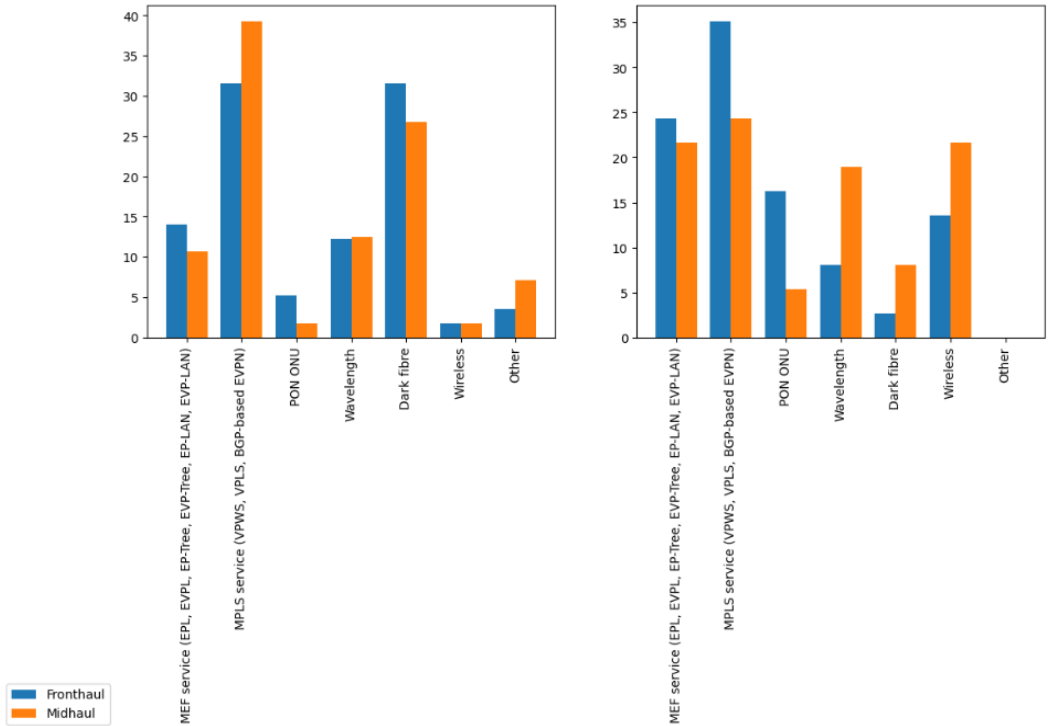

Fig. 9. Choice of fronthaul and midhaul technology for macro cell sites (numbers show percentage of sample size, NOG on left)

For fronthaul at small cell sites/fixed wireless access (FWA), what type of network service have you deployed most commonly?

A marked difference between the two profiles emerges here. Respondents from the NOG sample keep preferring MPLS service, but the incumbents lean towards EVCs, with MPLS second. Jointly, these account for 80% of choices expressed. NOG respondents' preferences are more evenly distributed, with dark fibre accounting for 27%. Technically, this is a sound decision; fronthaul is latency-sensitive and dark fibre supports the broadest range of applications, including the URLLC (ultra-reliable, low-latency communications) swathe of the 5G application space. EVCs and wavelength services are third and fourth most common. Wavelength service consists of purchase of one (or more) wavelengths on a CSP's CWDM / DWDM (coarse / dense wavelength division multiplexing) system. If the CSP purchasing the service demands that the wavelength not pass through any active device but only through passive devices, such as multiplexers/demultiplexers (mux/demux) and ROADMs, then the latency of this service should be no worse than that of a dark fibre with equal lightpath length. With MPLS and EVC, latency constraints are more severe and CSPs' choices in this regard indicate poor interest in provision of URLLC applications.

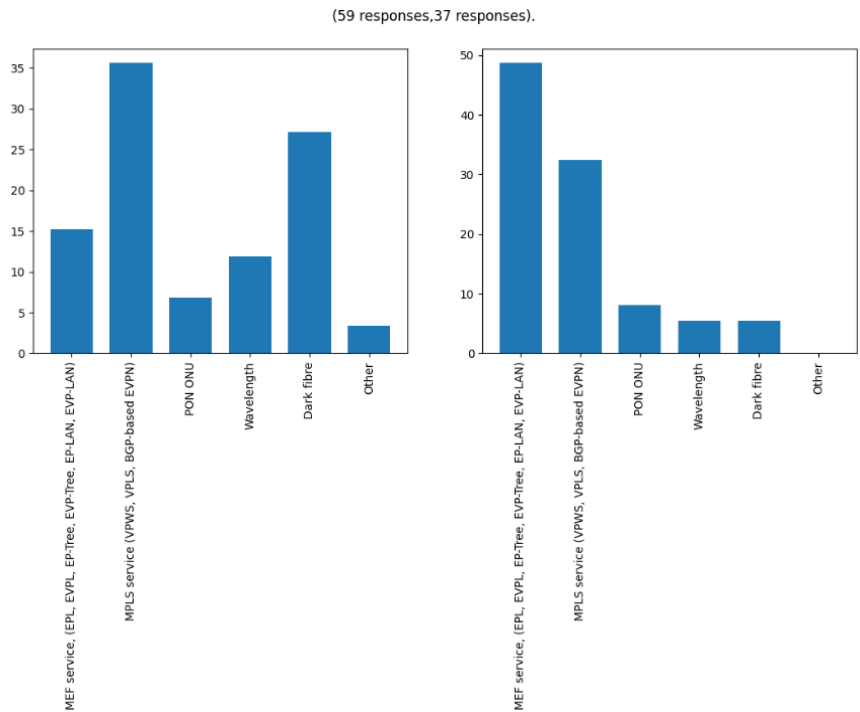

Fig. 10. Choice of fronthaul technology for small cell sites and FWA (numbers show percentage of sample size, NOG on left)

In your role as a carrier (if applicable), have you deployed disaggregated cell-site gateways (DCSGs)?

In essence, the DCSG's key attribute is that it enables separation of hardware and software on the cell-site device deployed for front-/mid-/back-haul of traffic from the cell site. Therefore, DCSGs support CSPs' need for leverage over their traditional suppliers of cell-site routing functionality. These devices are typically equipped with Ethernet or Ethernet + Synchronous Ethernet at layer 2 and layer 1, and IP/MPLS above layer 2. DCSGs are particularly relevant to CSPs who own or support other CSPs' radio access networks. NOG sample respondents show little inclination to deploy DCSGs, while the incumbents are heavily inclined to do so. This, too, may reflect the average size of individual respondents from the two samples.

*4.3 Aggregation architecture: layer 2 and higher*

At present, which form of layer 2 (or greater) aggregation of customer traffic from access node (V reference point) to service edges dominates?

Seamless MPLS refers to the facility to establish label-switched paths (LSPs) across segment demarcation points and CSP demarcation points. This is predicated upon the ability to exchange MPLS labels without the support of a domain-encompassing IGP. IGPs have two significant scaling impedances: one is the size of the IGP domain (the larger, the greater the scope for transient conditions) and the other is the autonomy of individual CSPs. The ability to exchange labels is implemented in BGP-LU [31].

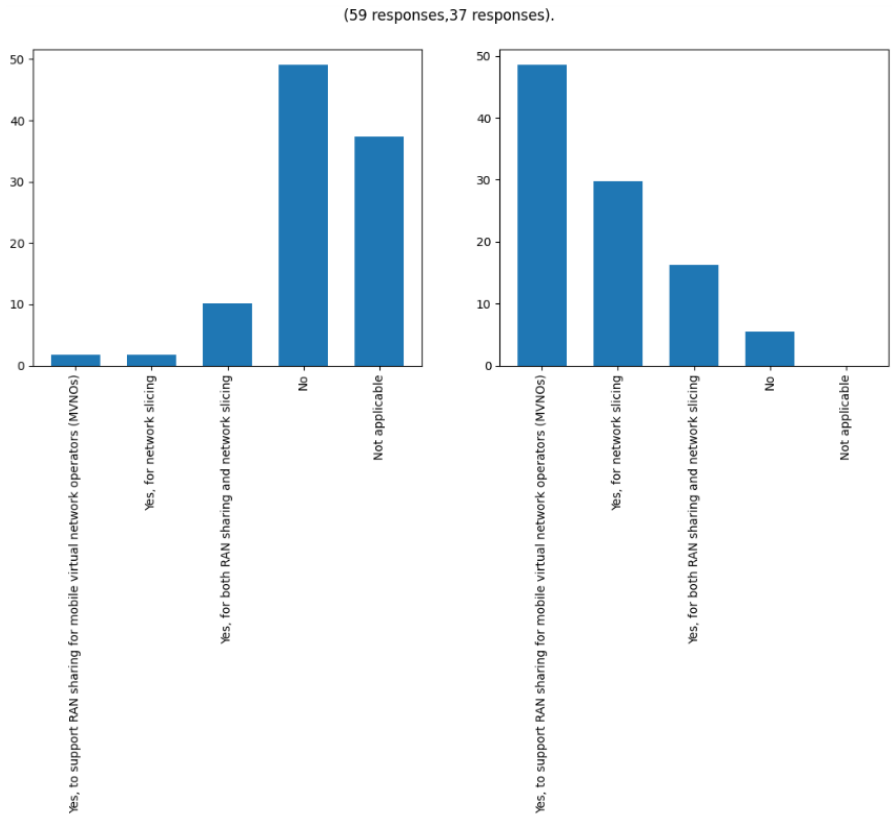

**Fig. 11.** Decisions of deployment of DCSGs (numbers show percentage of sample size, NOG on left)

Fig. 12 shows that seamless MPLS is the preferred aggregation technology among NOG respondents, with provider bridging [32] second and about half as popular. On the other hand, provider bridging [32] is more popular among incumbents, with seamless MPLS second. These choices are discussed in the analysis. Fig. 13 shows that respondents' choices do not vary significantly as regards what they would do now and in future implementations (follow-up question was: For aggregation of customer traffic from access node (V reference point) to service edges, which form would you tend to prefer for current and future deployments?).

Do you support the Ethernet Service Layer between the U1 and A10 reference points?

The MEF defines a three-layer Carrier Ethernet architecture (see [33, Fig. 2]), which focuses upon the Ethernet Services Layer (ETH) as the homogenizing transport across all customer access sites. Both samples clearly assert the necessity to support customer edge (CE) – PE connectivity over the Ethernet Services Layer (Fig. 14).

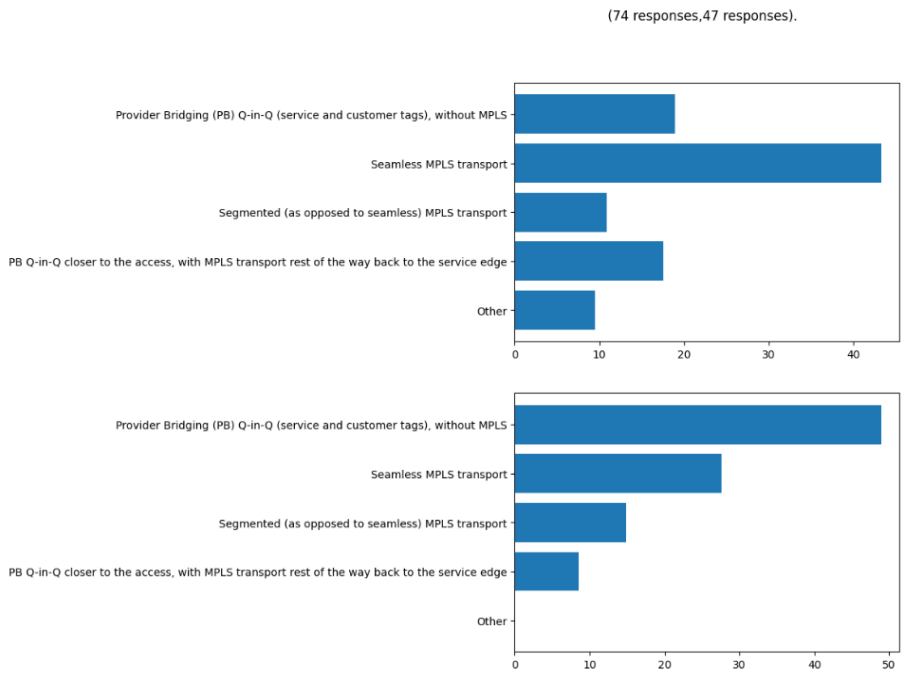

**Fig. 12. Layer 2 or greater aggregation AN – service edge (numbers show percentage of sample size, NOG on top)**

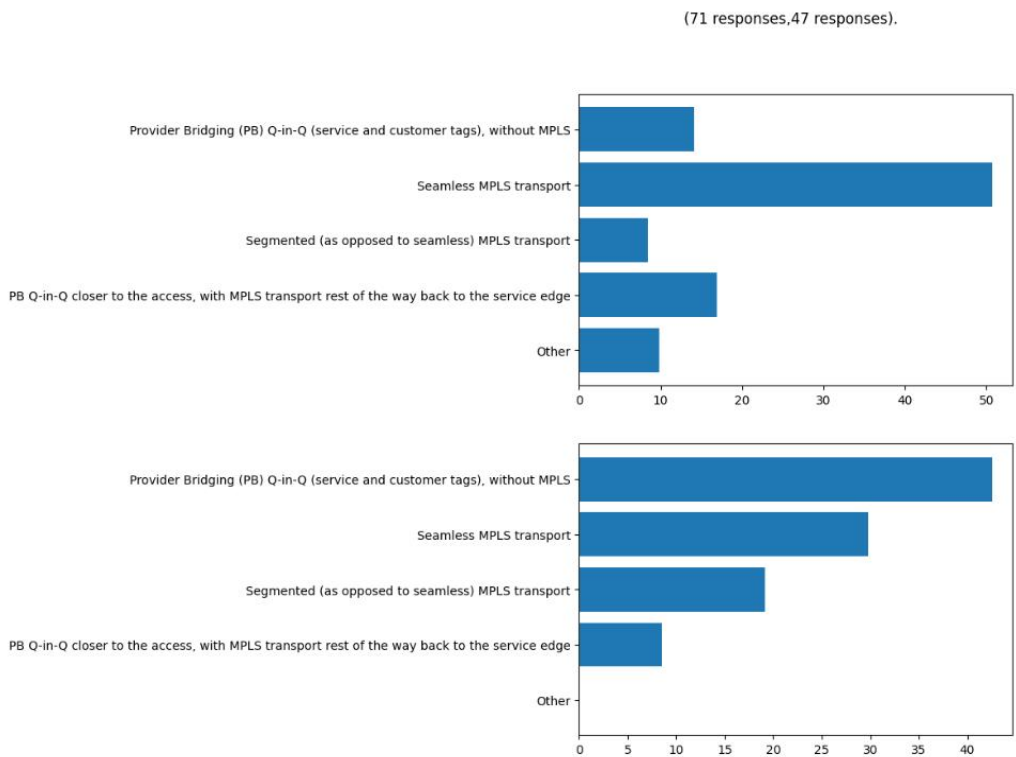

**Fig. 13. Future-oriented L2 or greater aggregation AN – service edge (numbers show percentage of sample size, NOG on top)**

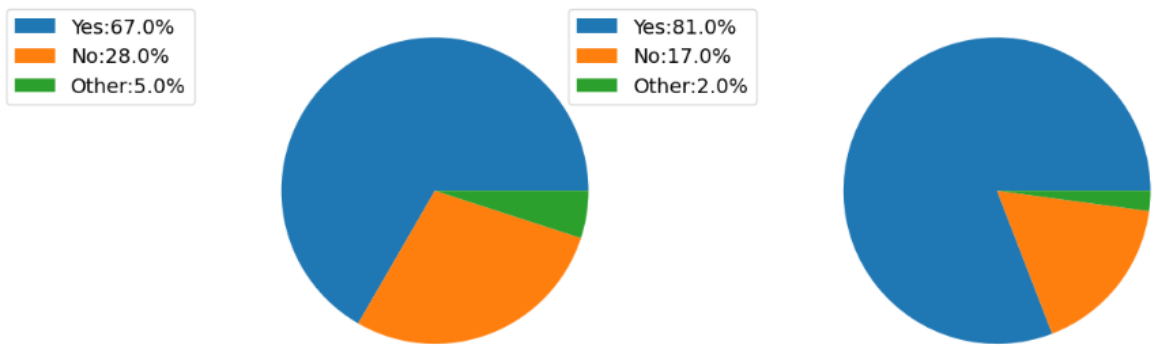

Fig. 14.  Support for Ethernet Services Layer from CE to PE (numbers show percentage of sample size, NOG on left)

If you answered yes to the previous question, is the Ethernet Service Layer your preferred means of layer 2 aggregation?

This question was placed with the support of a diagram from BBF TR-145 [34, Fig. 5]. It asks whether the CSP prefers offering the Ethernet Service Layer to customers (note that ETH is bounded by the UNI at one end) over other L2 aggregation technologies.

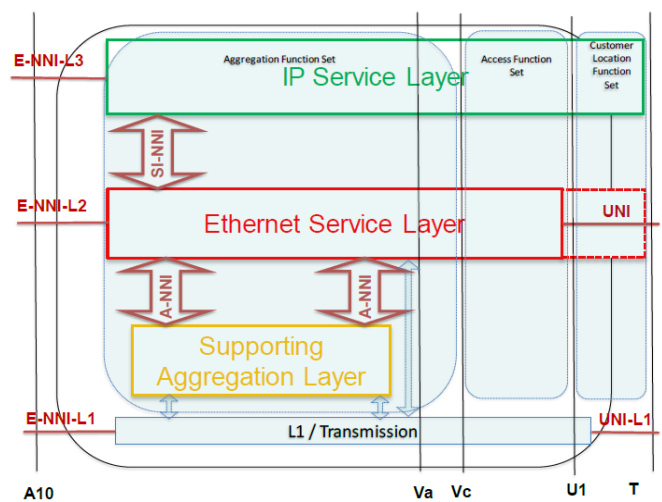

Fig. 15.  BBF TR-145's detailed reference model of multi-service broadband network (MSBN) [34, Fig. 5]

Fig. 16 suggests that, while the layer 2 Ethernet frame is not the universal transport mechanism, it is nonetheless a dominant one. This can be understood using two interpretive keys.

1.  The Ethernet frame is universally adopted in enterprise local area networks.
2.  Carrier networks are increasingly more equipped with means to implement Ethernet transport, whether it be VPWS (virtual private wire service), VPLS (virtual private LAN service), or the more recent Ethernet VPN (EVPN, [35]).

These two conditions, along with this result, suggest that ETH is on its way to universality as customer-facing L2 aggregation.

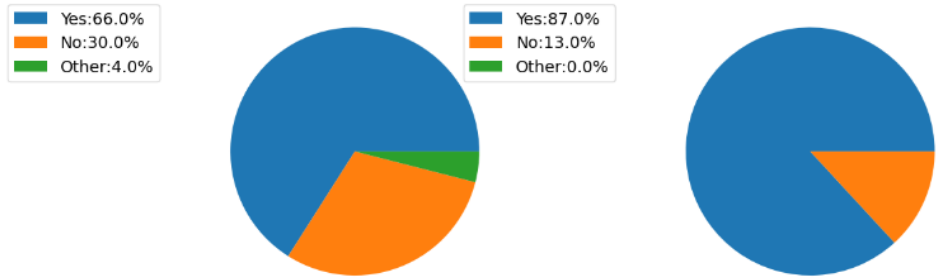

**Fig. 16. Preference for the Ethernet Service Layer as the means of aggregation (numbers show percentage of sample size, NOG on left)**

*4.4 Aggregation architecture: layer 1 and layer 0*

The following statements describe motivation for migration towards transport systems with integrated DWDM pluggable optics (and away from separate transponder/muxponder devices) and open optical line systems (and away from proprietary systems). For each motivation stated below, choose one response that best describes your opinion on its relevance as a motive for migration.

Respondents were asked to rate a series of motives (one motive per row) for moving:

1. away from DWDM systems that use separate transponder and muxponder units, and towards DWDM systems that use pluggable transceivers generating ITU-T – compliant wavelengths (coloured pluggables);
2. away from proprietary, book-ended optical line systems (OLSs), towards open OLSs.

The grid of choices is presented first (in tabular form), followed by a brief explanation of the options, and the results.

Table III    MOTIVES FOR CHANGE IN OPTICAL NETWORKS AND THEIR RELEVANCE

|  | 1[a] | 2[b] | 3[c] | 4[d] |
|---|---|---|---|---|
| DWDM optics can now be packed into switching and routing infrastructure face plates with the same density as client (grey) optics. |  |  |  |  |
| A line card can now carry a mix of grey optics and DWDM optics |  |  |  |  |
| 400ZR and 400ZR+ standardize the physical layer for metro area networks. |  |  |  |  |
| Open line systems facilitate use of interoperable pluggable DWDM transceivers. |  |  |  |  |
| Open line systems facilitate integration with existing management platforms. |  |  |  |  |

[a.] Mostly irrelevant
[b.] Somewhat irrelevant
[c.] Somewhat relevant
[d.] Highly relevant

The suggestions seek clarity on relevance (as a driving motive), as follows:

1. **Density of DWDM pluggables**, in terms of the number of pluggable transceivers that fit into a single rack unit (RU). This characteristic is one of the four objectives for improvement of pluggables, i.e., cost, space (density), reach and power consumption. The latest digital, coherent, optical transceivers are now manufactured in the QSFP-DD and OSFP packages, both of which enable the packing of 36 transceivers into 1 RU.

2. **Mix grey and coloured transceivers on a line card**: this facility supports flexible upgrades from transponder-based optical networks to integrated-pluggables optical networks.
3. **Standardization** – is this an important factor in moving towards pluggables and open OLS?
4. **Open OLS is expected to support interoperable pluggable DWDM transceivers** from different vendors at both ends of the link; is this an important driver towards open OLS?
5. Open OLS is expected to include open APIs – and therefore **integrate into the CSP's choice of vendor of network management system (NMS)**; is this an important driver towards open OLS?

The results are presented as clustered bar charts. Each colour represents a specific motive, as shown in the legend below the figure. The charts are most useful if the (Fisher-Pearson coefficient of) *skewness* of the individual sets of bars is considered. Take DWDM pluggable density (blue bars); this chart leans the most heavily towards the right for the NOG sample. Therefore, overall, NOG respondents consider this to be the most enticing motive to migrate towards integrated pluggables, i.e., that these pluggables can now be densely packed into router faceplates. On the opposite end of the skew range, standardization is the least enticing for the NOG sample. Table IV shows the order of relevance for both samples, basing on skewness.

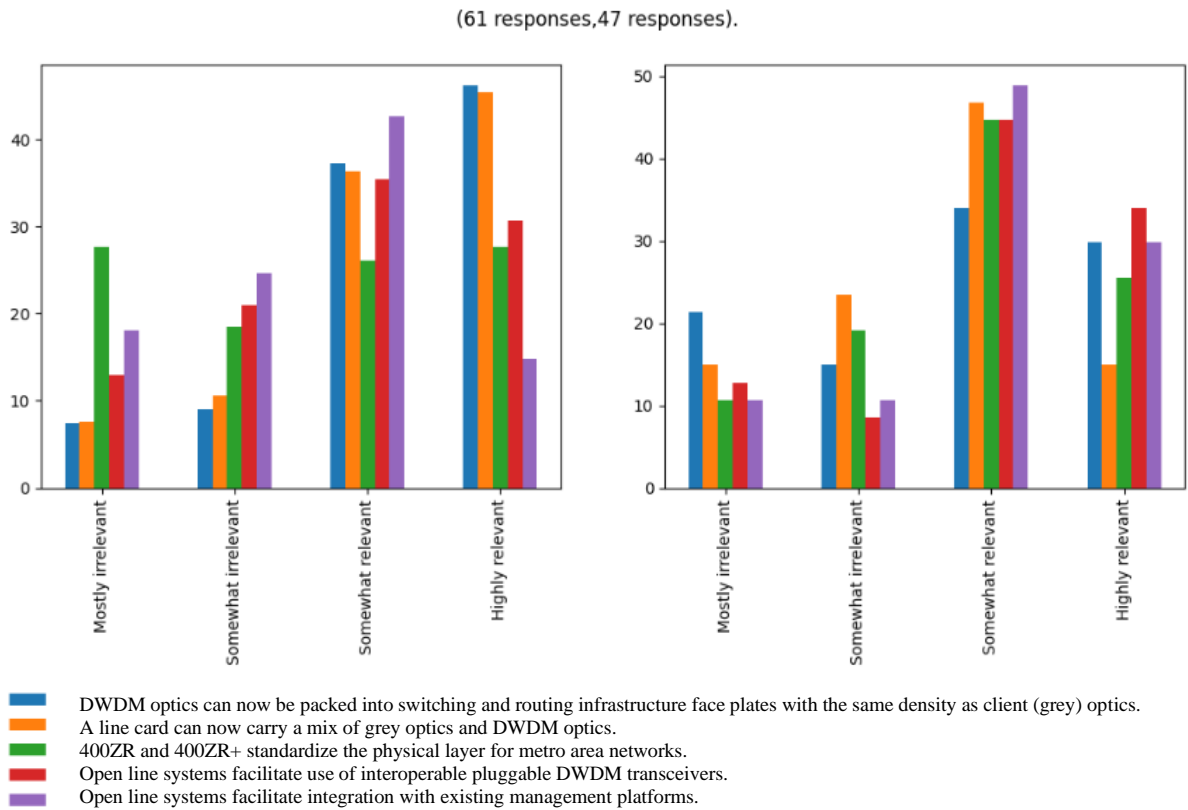

Fig. 17. Motives for migrations towards pluggable coloured transceivers and open OLS (numbers show percentage of sample size, NOG on left)

| Table IV | RELEVANCE OF SUGGESTED MOTIVES, IN DESCENDING ORDER |
|---|---|
| **NOG sample** | **SGA sample** |
| DWDM optics density | OLSs & interoperable pluggable DWDM transceivers |
| Line card can mix grey and coloured transceivers | OLSs & management platforms |
| OLSs & interoperable pluggable DWDM transceivers | 400ZR and 400ZR+ standardize MAN physical layer |
| OLSs & management platforms | DWDM optics density |
| 400ZR and 400ZR+ standardize MAN physical layer | Line card can mix grey and coloured transceivers |

XR optics enable a new point-to-multipoint network architecture. Do you plan to deploy this technology in your metro aggregation network?

"XR optics" is the catch-phrase for identification of an optical network technology that divides a wave band into a number of sub-bands, and supports aggregation of these sub-bands to fit different traffic distributions. This support is complemented by the facility to aggregate the sub-bands onto a single transceiver, thereby achieving a point-to-multipoint relationship between upstream and downstream transceivers.

The polarized positions of the two samples – the NOG respondents broadly do not plan to deploy, while all but three of forty-seven SGA respondents are in some stage of engagement – again seems to suggest a division along scale of operation. That incumbents have greater traffic volume and therefore greater scope for exploiting the technology's economies of scale, is reasonable and seems to emerge from the case studies (described later).

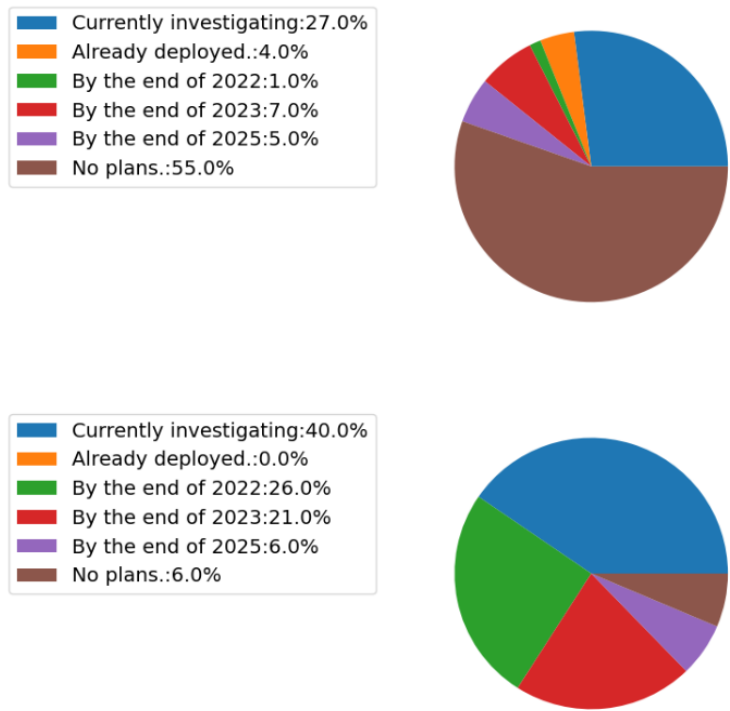

Fig. 18.   Plans to deploy XR technology (numbers show percentage of sample size, NOG on top)

Claim: "XR optics point-to-multipoint network architecture will replace all other network architectures in metro aggregation".

The purpose of this question was to gather feedback from CSPs about a technology that is relevant to two (2(a) and 2(b)) of the three pressures in the second major axis of the development framework.

NOG sample respondents are largely dismissive: counting lack of consideration, and lack of agreement on its potential to replace all other network architectures, just over two-thirds of respondents do not consider it to be a dominant technology. This fraction of the sample reflects the 55% that have no plans to deploy the technology.

SGA sample respondents, on the other hand, are overwhelmingly optimistic about the technology's future dominance. Almost all (95%) more ("fully agree") or less ("somewhat agree") think that XR optics' aggregation economics and granularity will displace other optical network aggregation technologies. On the basis of incumbents' scale of operations, this result suggests that the technology's implications must weigh heavily on the selection of scenarios for which to develop an implementational model.

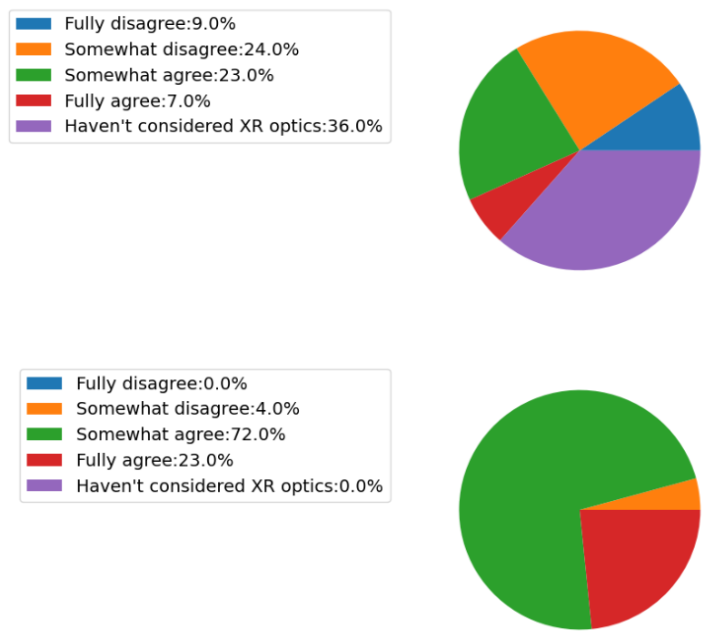

**Fig. 19.** Opinions on XR optics' prospect of dominating metro-aggregation (numbers show percentage of sample size, NOG on top)

Claim: "Existing OTN aggregation will stay in my network but I won't choose OTN for any expansion of my aggregation network".

Optical transport network's (OTN's) role in aggregation has been debated by a panel of CSPs and their vendors during an online symposium ([36, N. @60:25], and again at [36, N. @81:32]). An OTN equipment vendor identified it as a technology that can meet URLLC's low-latency requirement, but a CSP identified FlexE[37] for that role, and lamented OTN's cost: "*from the perspective of cost … definitely a no-brainer*". This anecdote is symptomatic of a broader debate over whether OTN is relevant to packet networking. OTN vendors cite physical separation (in separate OTN frames) as an advantage over purely packet networks, and cite higher efficiency (better utilization) over wavelength services. On the other side of the debate, cost, simplicity and (lower-cost) alternatives such as FlexE are cited as reasons for avoidance. However, even detractors see OTN's role in mid- and long-haul networks. There is a clear need for feedback from CSPs on their intentions with regard to OTN, and it was solicited by this question.

NOG sample respondents are inclined towards dropping OTN from their plans for the future of aggregation: 58% vs 42%. SGA sample respondents are **heavily** inclined towards dropping OTN. The overall verdict is that OTN's participation in the set of metro-area aggregation technologies is of secondary importance. The difference between the two samples might be attributed to NOG respondents' failure to respond to a question which they saw as irrelevant to them, as they do not operate OTN. Otherwise, the two distributions might be much closer to one another.

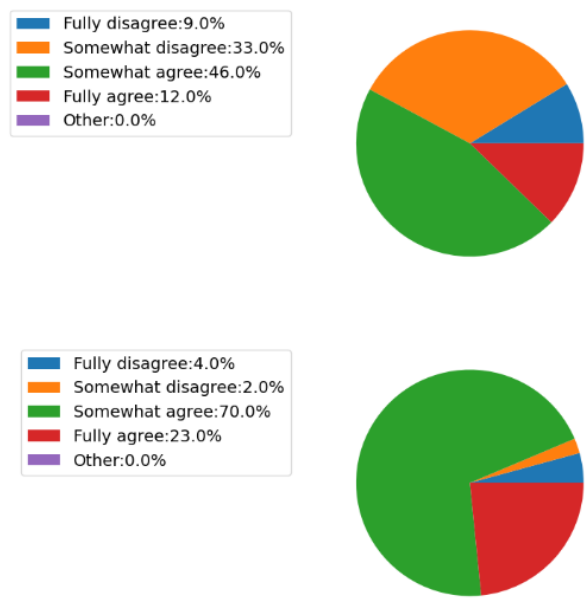

**Fig. 20.** Prospects for OTN's future in aggregation (numbers show percentage of sample size, NOG on top)

If you chose "somewhat agree" or "fully agree" that OTN won't be included in expansion of your aggregation network, please indicate the reasons driving your choice.

This question follows for those who are inclined to move away from OTN. Fig. 21 shows that cost is the primary driver for both respondent samples. The NOG sample's respondents included three explicatory comments: "*cheap fiber availability vs expensive OTN*", "*it is another management layer causing complexity*" and "*No Point doing TDM in today's world*". Apart from affirming cost, these comments indicate that simplicity would have been a good addition to the list of options offered (although: a catch-all option to suggest a reason was offered).

Response turnout was low on the NOG side. This too might be attributed to the same reason as that suggested to explain the previous question's turnout, i.e., lack of interest in a technology irrelevant to one's own technology set.

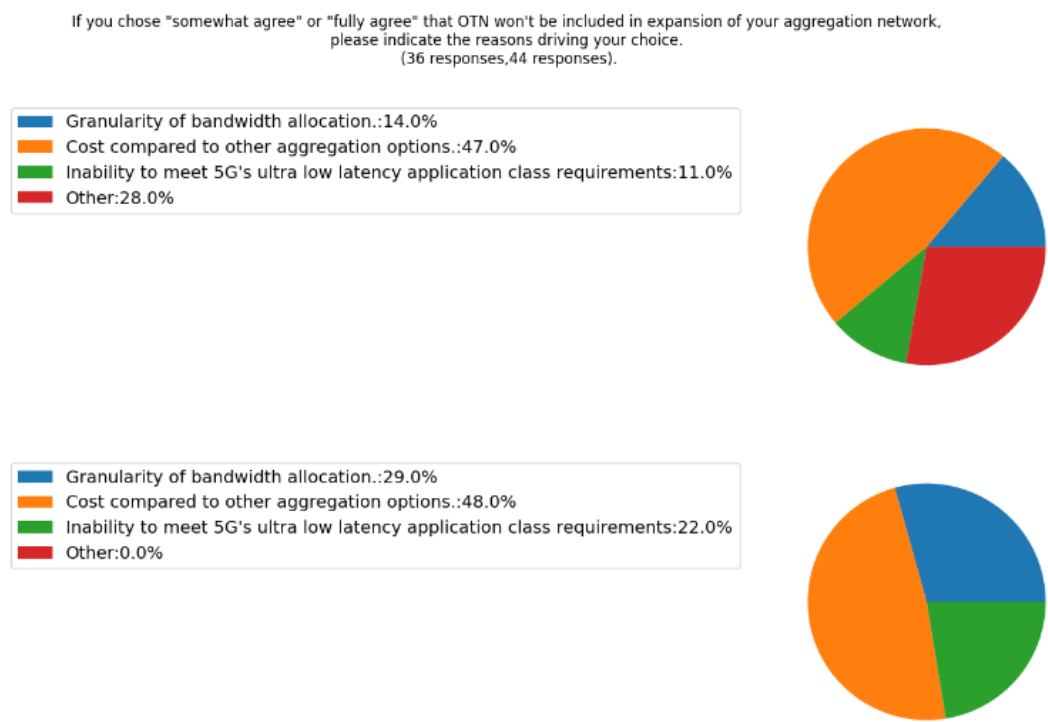

**Fig. 21.** Motives for dropping OTN from aggregation (numbers show percentage of sample size, NOG on top)

Claim: "Packet-based networks that share link capacities using soft slicing and/or hard slicing will fully displace OTN from metro area networks. The exception is in data-centre interconnect, where capacity allocations are stable."

This question was intended to assess respondents' inclination towards packet-based networks as an alternative to OTN. The basis for the exception for data-centre interconnect (DCI) comes from [38]; this source casts doubts on OTN's prospects for dominance, but identifies it as a means to groom client traffic (crossing datacentres, hence DCI) into optical payload units (OPUs) at varying bit rates. Fig. 22 shows the results: 64% of the NOG sample and 89% of the SGA sample lean towards a fully packet-based network. One respondent in the SGA sample chose the "other" catch-all and wrote "not sure". When this is compared with the results of the question about OTN's future prospects in metro-aggregation, it can be seen that more NOG respondents have a long-term vision of MANs without OTN, than those who favoured dropping the technology from the aggregation set. Moreover, once again, if the unresponsive subset of the NOG sample is considered as in tacit agreement, then the NOG sample would include 72% who are inclined towards purely packet-based networks, without OTN.

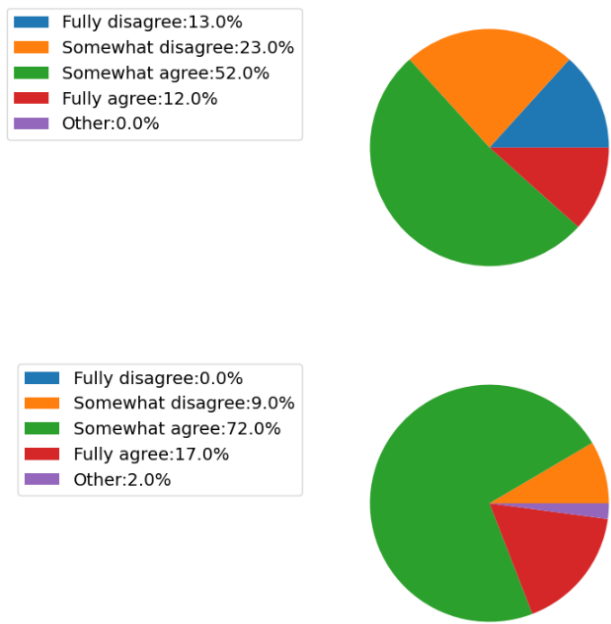

**Fig. 22.** Packet-based networks will displace OTN from MANs, except in datacentre interconnect (numbers show percentage of sample size, NOG on top)

*4.5 Transport network architecture: stacking layers, from 0 to 3*

The objective here is to detach from the details and attempt to acquire an understanding of trends in transport network architecture.

Which of the following best describes your current dominant form of metro-aggregation?

This question seeks to elicit an understanding of which technologies are occupying the stack of layers in transport, from layer 0 up to 3. Here, the qualitative analysis must be pre-empted, as otherwise the NOG sample's distribution cannot be interpreted correctly. Suffice it to state, for now, that "routed optical networks over Ethernet without ROADMs"[23] is taken to refer to the practice of drawing fibre up to the router chassis or shelf, and plugging it in to transceivers capable of meeting the optical link's budget – if possible, without line amplifiers.

With this proviso, the results (Fig. 23) seem to be very much in accordance with the broad divide between newer entrants (NOG) and incumbents (SGA). NOG respondents prefer the shallowest, least recursive stack of layers, and few of them are still using SDH/SONET. Their second preference – essentially the addition of coloured pluggable transceivers and ROADMs to support optical bypass of a switching node – is a distant second. It also emerges as the SGA sample's respondents first choice. For the SGA sample, OTN is broadly deployed as a sub-wavelength service layer for the carriage of Ethernet frames. OTN is far less popular among the NOG sample's respondents.

---

[23] A technically precise expression would have been "Ethernet over routed optical networks without ROADMs".

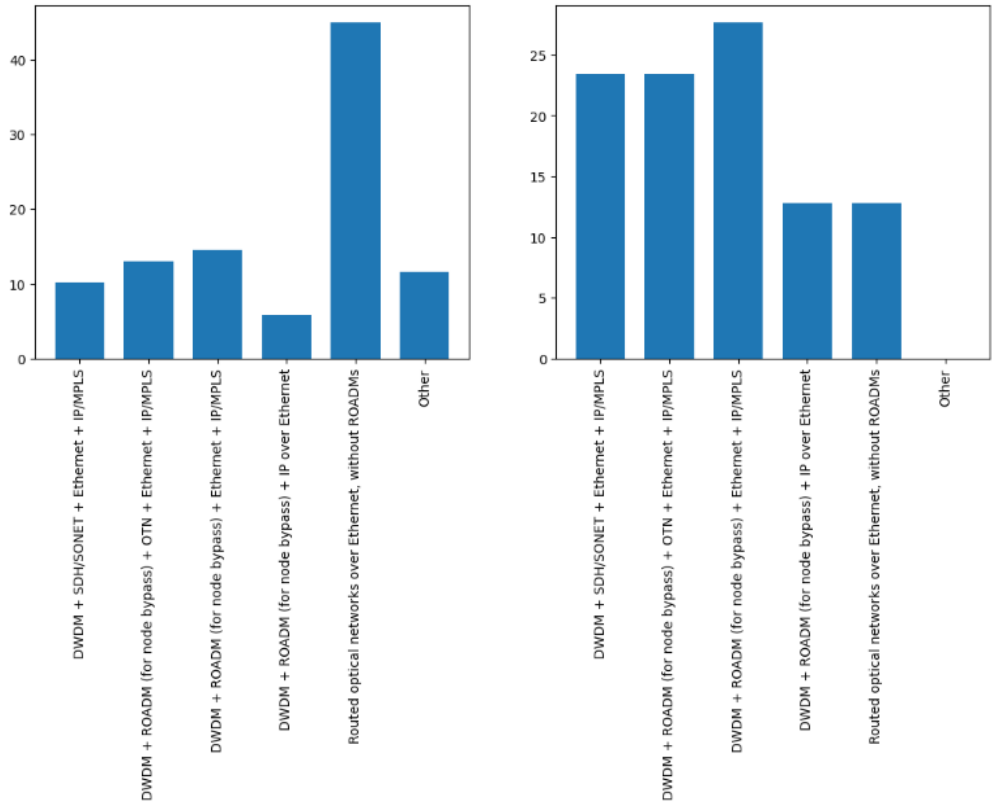

**Fig. 23.   Current dominant form of metro-aggregation (numbers show percentage of sample size, NOG on left)**

Which of the following best describes how you would deploy a greenfield form of metro-aggregation?

In order to understand which scenarios to model, it is necessary to solicit CSPs' plans for the transport network's future. The NOG sample's respondents' answers are consistent with previous responses: OTN's presence in the stack of layers comes in a distant circa 10% of respondents' choices. The most common choice (about 43%) is simply drawing fibre up to the switching node and plugging it in. A close second (about 33%) uses coloured pluggables and ROADMs. Note that MPLS is a part of the transport stack in every bar but the middle one, which accounts for less than 10% of the total.

SGA's sample's respondents' answers pose a difficulty. Notwithstanding the aversion claimed earlier to further development of the metro-aggregation span with OTN, here the technology is included in the most-commonly chosen stack of layers. When this issue was discussed with SGA's researchers, my attention was drawn to the uncertainty in respondents' previous answer: they "somewhat" agreed that they would not expand OTN. This leaves room for partial expansion, where specific customer requirements demand OTN's characteristics. A further, clarifying observation is that, overall, 36% would include OTN; the remaining 64% would not. When situated within the full context, the choices in favour of OTN retain overall consistency.  A remedial approach (for future efforts) might be to attempt to solicit clarification from respondents through hard-coded dependencies in questionnaires.

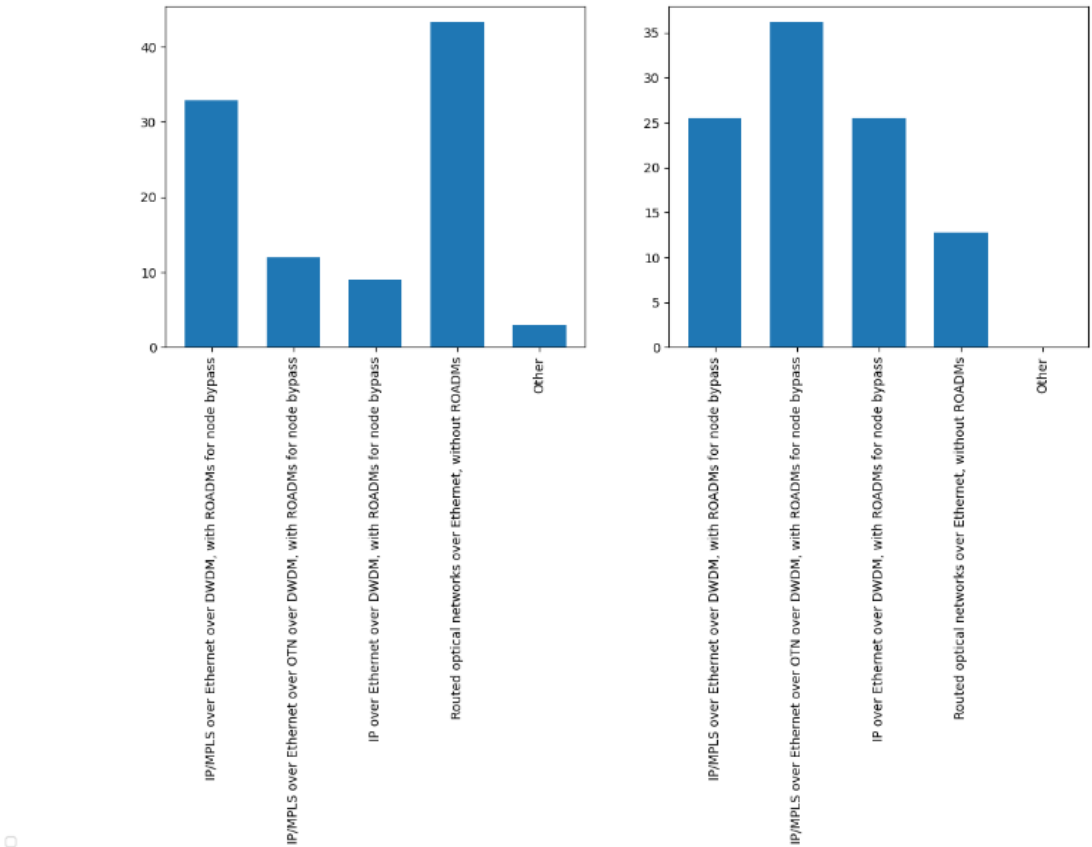

**Fig. 24.** Desired (greenfield) form of metro-aggregation (numbers show percentage of sample size, NOG on left)

Claim: "In the future, a mesh network will likely replace the metro-core ring at least in urban area with challenging capacity and resilience requirements."

The basis of this claim lies in [38], which states that "*the main driver for moving to a metro-core mesh is that it enables the introduction of IP-over-DWDM multilayer resilience schemes, with remarkable benefits in terms of enhanced reliability and optical interface reduction*"; in turn, this claim is rooted in [39]. Both papers come from sources that bridge the academia – industry divide; such claims, therefore, are weighty and merit investigation. The enhancement in reliability referred to in [38] (above) is enabled by fast-reroute (FRR), which can exploit loop-free alternate (LFA) and topology-independent loop-free alternate (TI-LFA). Since these reliability schemes are obtained through the higher-layer visibility at layer 3 than the rudimentary visibility available at layer 0, then the reduction in optical interfaces ensues. Rather than being limited to the addition of (costly) optical interfaces and employment of optical (network layer) protection switching (OPS), the network engineer is armed with LFA and TI-LFA as a means to achieving sub-50 ms switching in the event of link failure.

Fig. 25 shows an inclination towards agreement with the claim, for both samples. Two NOG respondents ("other" pie-slice) offered similar opinions: that such an architecture depends on (a) "geographies, costs and customer demand", (b) "the market, service area, and business case".

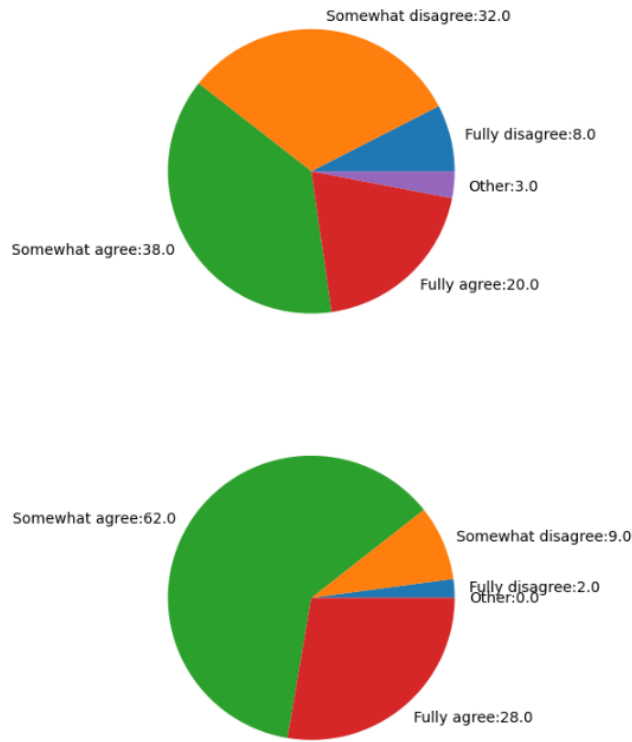

(66 responses, 47 responses).

**Fig. 25.** CSPs' inclination towards employing meshed over ringed metro-core nodes (numbers show percentage of sample size, NOG on top)

For greenfield metro-core deployment, how would you choose to implement an infrastructure based on DWDM optics?

For NOG sample respondents, the most common choice remains the simplest one: take (dark) fibre to the switching node, colourfully referred to as "point-and-shoot" (from router to router) by Arelion's (ex-Telia Carrier) representative in [40]: this is a reference to routes less than 40km in length and demanding less than 400Gb/s capacity aggregate across all wavelengths in a DWDM medium.

The NOG sample's responses (Fig. 26) are consistent with the collective mindset expressed in response to the metro-aggregation questions: keep it simple and only use ROADMs if you must. The "other" case shown in the left-side chart expressed "our top[o]logy – [ IP or Ethernet] over MPLS SR using BGP EVPN and coherent optics wherever required". This seems to qualify as the majority case, i.e., routed optical networks over Ethernet without ROADMs, and observes that the control plane uses Segment Routing. The SGA sample's responses are more evenly distributed; OTN is included in about 36% of responses and excluded from 64%. This is consistent with previous choices on the metro-aggregation stack.

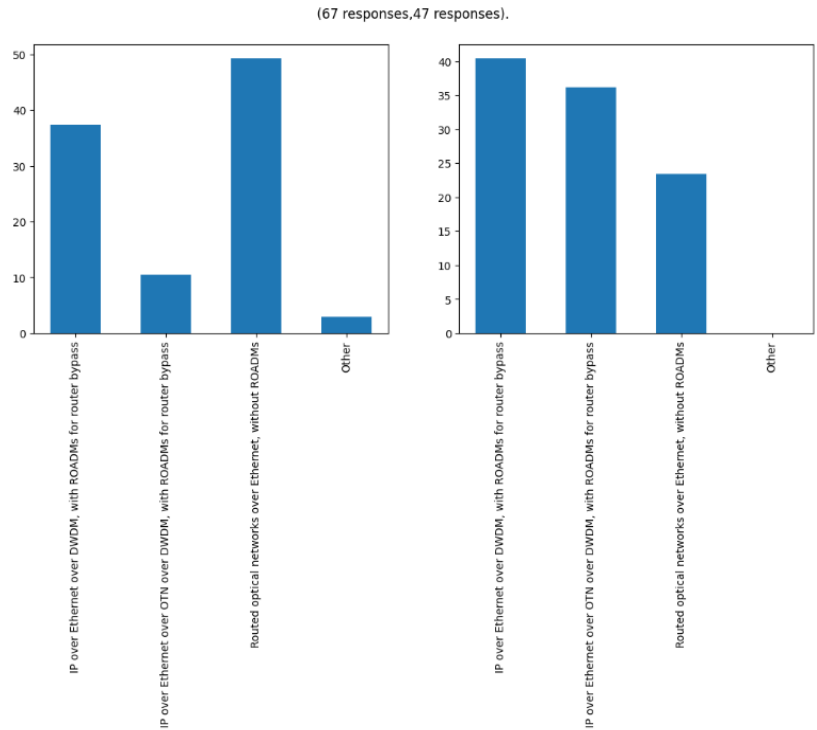

**Fig. 26.** Desired (greenfield) form of metro-aggregation (numbers show percentage of sample size, NOG on left)

*4.6 Service edge*

The final series of questions explored CSPs' thoughts about service edge development. In particular, thoughts on locating the edge closer to the customer were sought.

(A) Do you plan to deploy remote access nodes (Option 0) to enable MEC services? (B) Service edge locations (BBF TR-178): which do you currently employ for {Internet/Video} Broadband Network Gateway?

The two questions (in this sub-sub-section's title) complement one another and serve as an investigation of the relationship between CSPs' propensity to set up deep (close to the customer) service edges, and their disposition to deploy MEC nodes. The questions were accompanied by a graphic from BBF TR-178 (see Fig. 27, which is the same as Fig. 7, and reproduced here for convenience's sake).

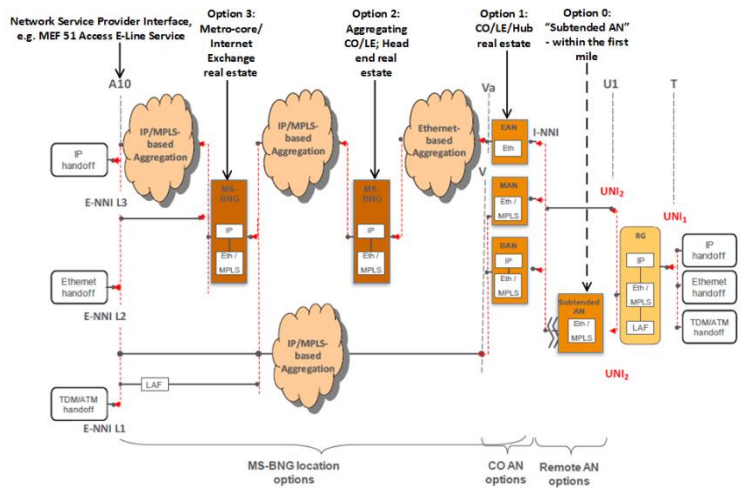

**Fig. 27.** General TR-178 architectural scheme, encompassing its targeted deployment scenarios [4, Fig. 2]

**Deployment of remote access nodes at the location indicated by Option 0**: Fig. 28 shows the distribution of responses. In the NOG sample:

- only 57 out of a total possible of 79 answered, and in this subset, 61% (35 out of 57) have no plans to deploy.
- Nine (9) already have deployed MEC nodes at the Option-0-location. However, out of these 9, only one has also deployed an Internet BNG at this location (this was determined through cross-reference within the nine respondents' answers).
- As regards those who installed an Internet BNG at the Option-0-location, none plan to deploy MEC nodes.

The above observations indicate that there seems to be no correlation between the two facilities. Similarly, only one NOG sample respondent has both MEC facilities and a video BNG at the Option-0-location.

As regards the SGA sample:

- all 50 participants responded.
- 12 (24%) claim to have deployed Option 0 MEC nodes, yet none have an Internet BNG at this site.

Given the similarity between the two samples in the lack of correlation between deployment of BNGs and MEC at the Option-0-location, one possible interpretation might be that the facilities that have to date housed BNGs, whether for Internet or video service, do not have sufficient infrastructural provision (power, cooling and security) to support MEC hardware. SGA sample respondents seem keen to move towards deep MEC nodes; indeed, as incumbents, they would have greater freedom to specialize personnel and re-purpose real estate.

One useful observation that emerges is the similarity of the two samples' distributions, and its match to the expectation that the mode of the distribution lies at Option 2 – i.e., the CO/LE.

(A) Claim: "Support for enhanced mobile broadband (eMBB) is improved by adding video BNGs closer to the end user.", and (B) Claim: "I would consider adding video BNGs closer to the end user to improve energy efficiency of video delivery."

These two questions were expected to draw a – more or less – affirmative collective response. Some NOG sample respondents declined to answer the question on improvement of support for eMBB, on the basis of their lack of support for a mobile network, or on their lack of support for video service. The NOG response to the claim for improved support for eMBB was further nuanced by some technical observations such as "Video is becoming more unicast, therefore it is a capacity planning equation that determines where video BNG is placed" and "Depends on distance of the BNG".

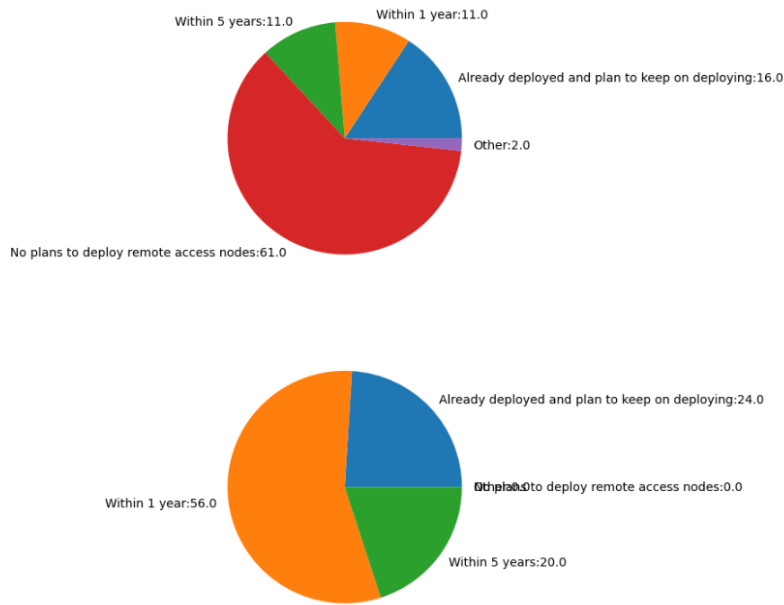

**Fig. 28.  CSPs' plans for deployment of remote access nodes deep into the access segment – Option 0 (numbers show percentage of sample size, NOG on top)**

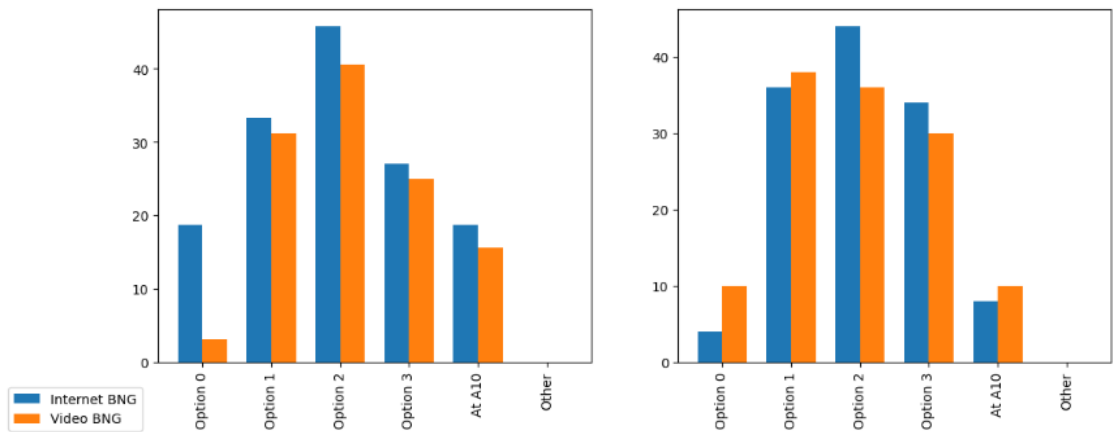

**Fig. 29.  Service edge locations supported by CSPs (numbers show percentage of sample size, NOG on left)**

Similar nuance was expressed by some NOG respondents on the issue of energy efficiency, e.g., "energy efficiency needs to be considered in its totality not only on the inte[r]faces.  Typically, the best energy efficiency is obtained in the data centres". A general statement can be made that the NOG sample reflects an expectation that both support for eMBB as well as its energy efficiency are improved by locating video BNGs closer to the end user. The SGA sample reflects a more emphatic expectation of improved support and energy efficiency. These results suggest that CSPs are favourably inclined towards deploying video BNGs.

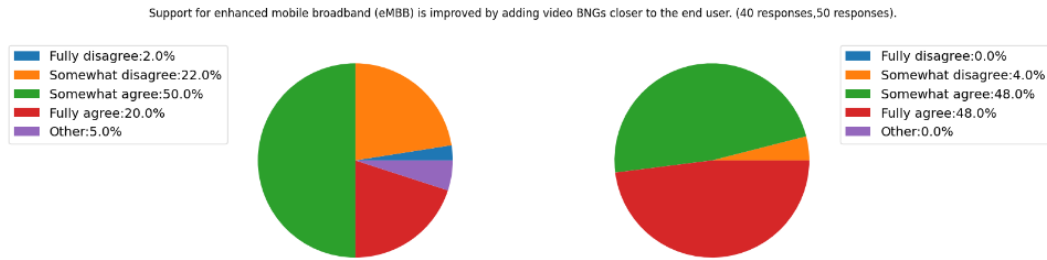

Fig. 30.  CSPs' understanding of whether support for eMBB is improved by adding video BNGs closer to the end user (numbers show percentage of sample size, NOG on the left)

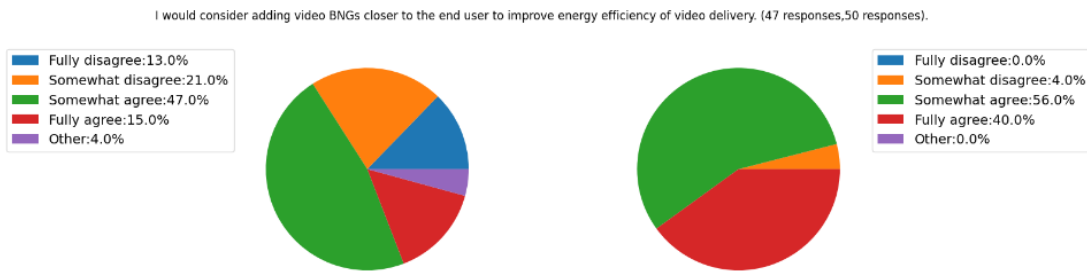

Fig. 31.  CSPs' understanding of whether energy efficiency is improved by adding video BNGs closer to the end user (numbers show percentage of sample size, NOG on top)

## 5. Qualitative survey results

The two broad objectives have been identified as discussion of the results' graphical summaries, and assessment of the questionnaire's questions' objective clarity. We have refrained from further qualification of the objective (e.g., it might seem plausible to express the objective as a desire to learn about reviewers' opinions on data accuracy), because the sources (interviews, emails, reports) at the foundation of qualitative surveys have the potential to supply raw data that exceeds, or at least encompasses, narrower objectives. All the resources mined for this qualitative survey are available online. The resources have already been referred to earlier in this work. Table V collects references to, and descriptions of the resources under one structure, for convenience's sake.

Table V  Resources collected during data stage of Qualitative Survey

| Resource description | Ref. # | URL |
|---|---|---|
| Face-to-face interview: anonymous_1[24] | | |
| Face-to-face interview: Ovidiu-Mădălin Roșet[25] | [22] | https://data.mendeley.com/public-files/datasets/k72dntnfgv/files/e3bafe8e-494a-4bcb-8b78-2d8db0525748/file_downloaded |
| Face-to-face interview: Haider Khalid[26] | [23] | https://data.mendeley.com/public-files/datasets/k72dntnfgv/files/093db0a7-c69f-40eb-95ff-ae1cf54684ba/file_downloaded |
| E-mail thread: Mark Tinka[27] | [28] | https://data.mendeley.com/public-files/datasets/k72dntnfgv/files/fba419c4-45c6-4332-917a-24babad0935b/file_downloaded |
| E-mail thread: Philip Smith[28] | [18] | https://data.mendeley.com/public-files/datasets/k72dntnfgv/files/a0a4bc7f-db2e-45c9-8920-c0fd056a5fe6/file_downloaded |
| E-mail thread: | [27] | https://data.mendeley.com/public-files/datasets/k72dntnfgv/files/a99a03fc-0e18-40f9-a45f-3ca4bdcec931/file_downloaded |

---

[24] We have not been given permission to publish the recording of anonymous_1's interview.
[25] https://www.linkedin.com/in/ovi12/
[26] https://www.linkedin.com/in/haider-khalid/
[27] https://www.linkedin.com/in/mark-tinka-5b03055/
[28] https://www.linkedin.com/in/philip-smith-154502/

| | | |
|---|---|---|
| Daniel King[29] | | |
| E-mail thread: anonymous_2 | [12] | https://data.mendeley.com/public-files/datasets/k72dntnfgv/files/dc3a3d2b-e188-4ca3-bd37-a15b34921802/file_downloaded |
| Written assessment: Haider Khalid[30] | [26] | https://data.mendeley.com/public-files/datasets/k72dntnfgv/files/079d7b49-4232-4709-bd88-1e624c71df00/file_downloaded |

In the interest of readability and concision, this section presents the results tersely, without dwelling lengthily on the process of transformation from data to results. An expanded form of the section can be found online[31], where excerpts of the raw data are cited as the bases for the results.

### 5.1 Face-to-face interview saliencies

Face-to-face interviews were conducted online, over Zoom, and recorded with the permission of the interviewees. In this sub-section, references to "I" and "me" in the course of references to the discussion, regard the interviewer, who was Etienne-Victor Depasquale. Duration ranged between 50 and 90 minutes. Reviewer anonymous_1 and Haider Khalid were recruited by SG Analytics, on the basis of the advice of reviewer anonymous_2, with which Mark Tinka concurred, to source interviewees directly involved with daily network operations. We recruited Ovidiu-Mădălin Roșeț directly, on the basis of the same advice and on the basis of personal experience of his technical networking skills. Both Haider and Ovidiu are CCIEs. Face-to-face discussions provided the opportunity to listen to narratives that described metro-aggregation and metro-core architectures. Such narratives address the degree of assurance which can be obtained about the integrity of the communication channel between questioner and respondent in the impersonal medium of the questionnaire. That is: since questions are text on paper, they are always, to some degree, subject to interpretation. The medium of a discussion reduces the subjectivity; the parties in a discussion have reasonable opportunity to solicit clarification in case of doubt. Indeed, the responses to the questionnaire *did* include the occasional unexpected interpretation. For example, one otherwise coherent respondent claimed, when asked about the location of the Internet BNG, that "BNGs have been dead for a long time! We use IP only, termination happens directly on the connected switch unless it is a wholesale circuit in which case it[']s dragged over to the handover". Here, it seems that the respondent was thinking about the service edge in terms of the PE router. This is indeed a possible interpretation: it is the "IP edge" that was referred to in [42]; what exactly has "been dead for a long time" is unclear.

Ovidiu-Mădălin Roșeț

On Layer 2 and higher metro-aggregation technologies

*"the old way of doing things"*

**Key takeaway:** Ethernet switches supporting spanning-tree protocol and basic VLANs were the cheap and effective way of aggregating from the access node upwards towards the metro core PE.

---

[29] https://www.linkedin.com/in/danielking/
[30] https://www.linkedin.com/in/haider-khalid/
[31] https://github.com/edepa/TrendsAndMotivationsInCSPMAN/blob/main/Section%205%20(expanded%20form).pdf

*"the intermedia[][te] step"*

I pressed for clarification on what the future of metro-aggregation looks like. Ovidiu's position as a system integrator enables him to broaden his vision beyond an extant portfolio of technologies, towards accommodation of a diverse set of customers, whether CSPs or otherwise. The answer branched into validation of NOG respondents' commonest choice of layer 2+[32] aggregation (seamless MPLS), and then branched again towards a welcome insight into how large routing domains can be constructed.

**Key takeaway:** Seamless MPLS is enabling "IPfication" of the network, wherein the CSP is able to bring IP connectivity into the access network. Moreover, seamless MPLS, through BGP-LU, supports end-to-end LSPs in large networks.

*The future*

My line of questioning moved towards provision of MEF-compliant ETH layer services, which, operators claim to support in their overwhelming majority (see Fig. 14).

**Key takeaway:** EVPN, signalled with the support of Segment Routing in the control plane, is the successor that takes up the mantle of adopting MPLS in the data plane.

### On Layer 0, 1 metro-aggregation technologies

The interview proceeded towards discussion of results of asking CSPs about their motivation for migrations towards transport systems with integrated DWDM pluggable (away from separate transponders and muxponders), and towards open OLS.

**Key takeaways:**

1. DWDM is new technology, unlikely to be widely deployed[33] because it requires new hardware.
    a. This helps to explain the dominance of routed optical networks.
2. Apart from operational simplicity, another significant advantage is reduction in OPEX, from saving electricity on separate transponder equipment.

### On technology stacks for metro-aggregation

It was time to address stacks of technology; this was the opportunity to ask about interpretation of "routed optical networks"; Ovidiu did not expect respondents to relate the term to Cisco. This, of course, corresponded to [Eduard Vasylenko's claim](#) about the lack of association between the term "routed optical networks" and Cisco's use thereof.

**Key takeaways:**

1. Routed optical networks are most likely interpreted as fibre drawn between routers.

---

[32] Layer 2+ is a somewhat loose term that is used here to refer to technologies that fit above OSI layer 1 but below OSI layer 3. In this work, the only significant members of the implied set are IEEE 802, 802.1, 802.3 and MPLS (RFC 3031).
[33] The likelihood referred to pertains to the need to amortize installed equipment. In the circumstance that equipment is not fully amortized, an operator may have to delay installation of newer technology.

2. It was reaffirmed that WDM may not need to be considered, given a glut of fibre.

Haider Khalid

On clarity of the questions in the questionnaire

*Overall*

As regards the questions, I asked "*whether you feel they were clear, whether they were open to interpretation, and whether there are things that really needed to be improved*". Haider's difficulty was limited to those questions regarding "transmission", which is out of his scope.

I then proceeded to enquire about the dissemination of knowledge about standardized reference points (T, U, U1, V, A10) among network engineers. While reference to these RPs was always accompanied by an explanatory graphic in the questionnaire, I wanted to understand obstacles to apprehension both for the survey's purpose and to form my own understanding of jargon familiar to the groups within my research's scope. Haider affirmed the use of UNI and NNI, but not the RPs, citing U1 and V specifically. These comments cast some doubt on the use of reference points among CSPs' technical personnel, and so I pursued this further, and referred to the TR-178 graphic (Fig. 7). "Access [edge]" and "provider edge" were correctly aligned with the V and A10 RPs. This was a good indication that this graphic was suitable as the basis of a number of questionnaire questions.

*On the interpretation of the term "routed optical networks"*

Haider recognized the term's proprietary nature, but suggested that Cisco's position as *"the go to vendor for everybody"* might compensate.

**Key takeaways:**

1. Reference points may not be widely known among CSPs.
2. Routed optical networks is not a term that has precise meaning.

On Layer 2 and higher metro-aggregation technologies

**Key takeaways:**

   1. When Haider's comments are combined with Ovidiu's and with the statistics, it emerges that the subset of Regional and Tier 1 operators which are active in the access and aggregation, are more likely to operate:
      a) IEEE 802.1Q-2022 aggregation all the way up to the metro core, and
      b) IP/MPLS switching in the metro core.

On Layer 0, 1 metro-aggregation technologies

Here, discussion regarded:

- results of the question asking CSPs to rank their motives for migration towards integrated DWDM pluggable and open OLS: Haider agreed that the facility to pack

- DWDM pluggables densely is an important motivator, as well as the importance of mixing of grey and coloured pluggables. For the NOG respondents, these were the motives that leant most and second most towards high relevance.
- the role which OTN would play: Haider indicated that SDH is still in use in his organization, but that it is to be replaced with DWDM all across the transmission. The reply was coherent. The organization involved is a large one, with several million subscribers. The need for return on investment in SONET/SDH technology would have supported its retention well past some form of collective realization in the sector of CSPs that the technology had been superseded.
- topology: I sought Haider's perspective on the results about the question whether meshes are likely to be more common in the future among metro-core nodes. On this reading, even access nodes aggregators (e.g., the Ethernet switches upstream of the V RP) are fully meshed in 70 – 80% of CSPs. This does agree with one anecdote which I can personally relate with regard to a local CSP. Rings, therefore, while apparently convenient, do not give CSPs the desired level of assurance on service availability.

**Key takeaways:**

1. While rings are commonly used as examples in literature (see, for example [38], [43, Ch. 17]), mesh interconnection of access nodes aggregators (for emphasis's sake: access nodes are devices like DSLAMs, CMTSs and OLTs, which aggregate subscriber lines) is at least equally likely.
2. Another example of SONET/SDH's removal from aggregation technology stacks, was given.

On technology stacks for metro-aggregation

**Key takeaways:**

1. Regional and Tier 1 CSPs active in the metro area are adopting IP/MPLS over Ethernet over an optical network comprising DWDM links that are optically switched using ROADMs.

Reviewer anonymous_1

On Layer 2 and higher metro-aggregation technologies

**Key takeaway:**

1. Further emphasis is made on what was observed earlier about aggregation in regional operators' networks:
    a. IEEE 802.1Q-2022 aggregation all the way up to the metro core, and
    b. IP/MPLS switching in the metro core.

On Layer 0, 1 metro-aggregation technologies and technology stacks for metro-aggregation

I addressed the motives for migrating towards integrated DWDM pluggables and about OTN being displaced by packet networks.

**Key takeaways:**

1. Return on investment is a key criterion in determination of rate of penetration of replacement technologies. SONET/SDH is particularly hard to displace in the "local PTTs[34]" – these are the CSPs of a smaller, regional scope, the descendants of what are colloquially referred to as the "Baby Bells", when AT&T was broken into Regional Bell Operating Companies.
2. It was reaffirmed that elimination of transponders is a primary determinant in migration to integrated DWDM pluggables.

*5.2 Written media: e-mails and reports*

The two broad objectives (discussion of the results' graphical summaries and assessment of the questionnaire's questions' objective clarity) were further pursued through written media. This approach facilitates reflection on both parties' sides, whilst lacking the immediacy obtained in face-to-face interviews. For convenience's sake, the portion of Table V pertinent to written media, is reproduced below as Table VI.

| Table VI | | WRITTEN RESOURCES COLLECTED DURING DATA STAGE OF QUALITATIVE SURVEY |
|---|---|---|
| E-mail thread: Mark Tinka[35] | [28] | https://data.mendeley.com/public-files/datasets/k72dntnfgv/files/fba419c4-45c6-4332-917a-24babad0935b/file_downloaded |
| E-mail thread: Philip Smith[36] | [18] | https://data.mendeley.com/public-files/datasets/k72dntnfgv/files/a0a4bc7f-db2e-45c9-8920-c0fd056a5fe6/file_downloaded |
| E-mail thread: Daniel King[37] | [27] | https://data.mendeley.com/public-files/datasets/k72dntnfgv/files/a99a03fc-0e18-40f9-a45f-3ca4bdcec931/file_downloaded |
| E-mail thread: anonymous_2 | [12] | https://data.mendeley.com/public-files/datasets/k72dntnfgv/files/dc3a3d2b-e188-4ca3-bd37-a15b34921802/file_downloaded |
| Written assessment: Haider Khalid[38] | [26] | https://data.mendeley.com/public-files/datasets/k72dntnfgv/files/079d7b49-4232-4709-bd88-1e624c71df00/file_downloaded |

Mark Tinka

Mark's support in crafting the questionnaire precluded any discussion on the questions' clarity. The results were within scope of discussion; a first reading drew some scepticism, due to the dominance of ADSL2+ in SGA's sample, both as the largest and fastest-growing access technology. The ensuing exchange with SGA's representatives accentuated the importance of balancing quantitative surveys with qualitative surveys. Therefrom, it emerged that with ***regional and Tier 1 CSPs*** dominating SGA's, and with further suggestion that these were indeed incumbents in their markets, the SGA sample includes CSPs who are still reaping returns from their investment.

---

[34] Postal, telegraph, and telephone service
[35] https://www.linkedin.com/in/mark-tinka-5b03055/
[36] https://www.linkedin.com/in/philip-smith-154502/
[37] https://www.linkedin.com/in/danielking/
[38] https://www.linkedin.com/in/haider-khalid/

**Key takeaways:**

1. Return on investment is a key criterion in determination of rate of penetration of replacement technologies. The observation here arose in the context of an access technology (copper-based ADSL2+).
2. The SGA sample's response is representative of incumbents with legacy infrastructure still being monetized.

Philip Smith

While Philip drew attention, as respondent anonymous_2 did, to uncertainty with the representativeness of results, he asserted that the results did not jar with his understanding of current metro area networks and developments thereof. Moreover, in the process of evaluating the credibility of the data, a fresh perspective on the two groups was offered: NOG respondents are likely to include those with a higher degree of autonomy in taking decisions than those from the large incumbents.

I also invoked Philip's support on the issue of interpretation of the problematic term "routed optical networks". In view of the significance of this technology stack's impact, I have deferred citing his contribution towards resolution of the issue, to the analysis.

**Key takeaways:**

1. Given the observed difference between the two samples, the results match expectations.

Daniel King

Limitations and ambiguities in the questionnaire were addressed directly. I also asked how well the results match his perception of trends. Given Daniel's background as witnessed by his participation in several RFCs[39], I also asked for his opinion on the widespread choice of seamless MPLS as a layer 2+ aggregation technology.

**Key takeaways:**

1. Multi-layer packet routers are facilitating delivery of differentiated end-to-end services – and these services' availability benefits from the smaller domains facilitated by seamless MPLS.
2. The results match expectations.
3. The questions are clear.

---

[39] https://datatracker.ietf.org/person/d.king@lancaster.ac.uk

In addition to the face-to-face interview, Haider accepted to dwell further on the content and wrote a brief report [26]. The key takeaways from the report are reproduced below.

**Key takeaways:**

1. UNI and NNI are easily recognizable terms; U1 and A10 are not.
2. In similar vein: PE is recognizable as the edge of access. This calls to mind the Stage 2 segmentation model [42].
3. The results match expectations. Haider cited the following as particular cases of agreement between results and expectations:
    a. distribution of deployment of access technologies;
    b. Ethernet as a "major layer 2 backhaul";
    c. MPLS as dominant switching technology, especially when core is taken into account, and
    d. location of video BNGs as close as possible to the end-user, to save bandwidth.
4. No ambiguities in the questions were detected (though the graphs presented to Haider for his analysis, were found lacking in clarity).
5. Limitations observed concerned the desire to extend into questions on SDN. Here, my defence is the same as that offered with regard to Daniel King's observation on scope of the quantitative survey.

## 6. Analysis

This section is divided into two parts. The first sub-section carries results' saliencies, obtained as a product of the quantitative data and discussions thereon. Structured discussions were held with named reviewers, while other field experts contributed to specific issues. The second sub-section presents a first set of scenarios that emerge as candidates for full implementational modelling.

*6.1 Saliencies emergent from quantitative and qualitative survey*

1) The most common technology stack: routed optical networks

The first *hint* at the interpretation of this popular choice of metro-aggregation technology stack which CSPs are deploying emerged from discussion with Ovidiu-Mădălin Roșeț [22, N. @38:23]. ***Clear evidence*** emerged from Philip Smith [18, N. See mail on June 27th 2023]. Philip emphasized that the interpretation is purely dark fibre plugged into routers, with the need for simplicity strongly driving this approach, notwithstanding vendors' offers enabling channelization, whether through wavelengths or timeslots. Moreover: re-visitation of an e-mail exchange held earlier with a small (1-10 metro areas, 1 – 100k subscribers) CSP's CTO [44] affirmed this understanding. I had asked where operators were gravitating in the tension between circuit- and packet-transport MAN technologies. I also asked about this CSP's use of an OTN. The reply affirmed the principle of simplicity, notably where short distances (less than 40 km, again) are involved: it is cheaper to deploy optical interfaces directly onto fibre, without OTN or DWDM, but longer distances between intermediate nodes change the bias in favour of consideration of such options. Therefore, it is reasonable to conclude that, unless qualified otherwise, the term "routed optical network" evokes the simplest operational architecture that supports packet switching, namely:

1. Layer 0: dark fibre
2. Layer 1: a pluggable transceiver, possibly using grey optics such as 10GBASE-ER, or possibly using coloured pluggables in anticipation of their future use.

Given the extensive adoption of Ethernet and MPLS, and the ubiquity of IP, it seems fair to extrapolate further to cover layers 2 – 3 with these technologies. One additional consideration is worthwhile. While discussing a related technical matter with Mark Tinka, my attention was drawn to the extent to which some CSPs take simplicity [45]: a number of operators, "religious about simplicity", run MPLS-free backbones and forward all of their traffic using IP. Indeed, one CSP, while answering the question about the technology stack, opted to answer with his own text (using the "other" catchall), and wrote "dark fibre + Ethernet + IP". Regardless of whether MPLS is employed to forward traffic in the data plane or not, the convergence of the various sources cited here strongly indicates the intended composition of the stack of layers implied by the common choice "routed optical networks".

2) Layer 2 and higher aggregation from the access node to the provider/service edge

In these layers of the technology stack , when all data – both from quantitative and qualitative survey – are processed, they cohere well and conclusions can be drawn on the state of current networking and the expectation for next-generation networking. Seamless MPLS was identified as [an intermediate step](#) and of current interest to CSPs. It is [strongly preferred by NOG respondents](#) as an aggregation technology, but Provider Bridging is most common among Tier 1 and regionals (SGA). Given the larger average subscriber base size of Tier 1s and regionals, it matches intuition to find that the rate of adoption of newer technology is greater among the group of CSPs (the NOG set) with smaller average subscriber base size. Therefore, on the basis of data collected from the two surveys, it can be observed that current layer 2 and higher aggregation, in descending order of deployed instances is the following.

1. Most deployed
    a. V to A10: IEEE 802.1Q-2022 Provider Bridging, or QinQ
    b. Beyond A10: MPLS
2. 2nd most deployed: Seamless MPLS, from V to beyond A10

Moreover, deployment of next-generation aggregation will be ordered as follows:

1. Most deployed: Seamless MPLS, from V to beyond A10
2. 2nd most deployed

       a. V to A10: IEEE 802.1Q-2022 Provider Bridging, or QinQ
       b. Beyond A10: MPLS
3. 3rd most deployed: Segment Routing controlled multi-domain MPLS

It is difficult to anticipate that SR-controlled multi-domain MPLS could easily overtake both other aggregation sets. This is due to the paradigmatic change represented by SDN, which requires cultural change within CSPs. Any further comment at this point would be little better than pure conjecture.

3) Distributed Access Architecture – a split along sample boundaries

By far the largest percentage (44.4%) of NOG respondents are not planning a distributed access architecture; in contrast, it is the smallest percentage by far (2.1%) of SGA respondents that are not planning a DAA. One respondent in the NOG sample helpfully opined that "[a] distributed access architecture is applicable in large metro areas. In small metro areas (in terms of geographical coverage) it is typical to concentrate equipment in a few PoPs.". If the smaller mean, median and mode of the NOG sample are indicative of the geographical coverage of CSPs in the sample, then the difference between the two samples might be explained on this ground.

The importance of DAA lies in its inherent infrastructural demand for a new RPI-N: that between the two actors identified in [46, p. 67]. Even if the Virtual Network Operator and the Infrastructure Provider are "subsumed into a single entity", a RP[40] must be identified for the interconnection within the designated real estate: the "NFVI-PoPs such as central offices, outside plant, specialized pods" [46, p. 16].

4) Correlation between DAA and virtualization; between DAA and customer-proximal MEC

65% of NOG respondents do not plan to fully virtualize; the same percentage does not plan to employ DAA for the majority of HHP. Moreover, 44.4% of NOG respondents have no plans at all for DAA. On the other hand: all SGA respondents plan to fully virtualize within 5 years or less, or have done so already, and the same percentage – 2% – does not plan to use DAA as that which does not plan to serve the majority of HHP with DAA. A similar correlation exists between DAA and Option 0 (see Fig. 7) MEC nodes. In more detail than is visible in the charts: of the 32 NOG respondents who are not planning any DAA, only 5 indicated that they are deploying MEC in locations close to the customer (option 0).

These two correlations seems to justify the observation carried in [46, p. 67], that "virtualising broadband access nodes can be exploited by the co-location of wireless access nodes in a common NFV platform framework … thereby improving the deployment economics and reducing the overall energy consumption in the combined solution." Admittedly: a wireless access node is not a DAA node, but it must be conceded that any extant wireless access node is a candidate for extension of its role into that of DAA node.

---

[40] An RPI-N is a specialized RP. If roles on either side of the RP are subsumed by one entity, then the scope for an RPI-N is replaced by that of an RP.

*6.2 Scenarios*

The number of aggregation and core scenarios is too large to handle well all at once. Rather, it seems more useful to use the common implementations as the bases upon which to support future modelling work. Of course, the surveying techniques used here are only part of the methodology (see [section 2 of this paper](#)), but they are a significant and valid part. In this sub-section, the results of surveying will be used to compile a set of aggregation and core scenarios for modelling.

1) Layer 2 and higher aggregation

A first scenario: PB, with G.8032 and TR-101 N:1 forwarding

With the larger subscriber-base CSPs (SGA sample), an initial reading of Fig. 13 shows a dominant Provider Bridging (Q-in-Q) component. Some limitations must be addressed before this common choice can be interpreted into a realizable implementation.

*Limitation: number of service deployments*

Used without MPLS, PB is limited to roughly 4000 different service deployments. Indeed, BBF TR-101 Issue 2 [10, p. 32], which describes Ethernet-based broadband aggregation states explicitly that "Aggregation Switches will only forward based on S-Tags" and requires the "uniqueness of a S-Tag across ANs". For example, if an enterprise customer were to purchase an attachment circuit (AC) connecting the CE switch (the UNI-C device) to the provider's UNI-N device in a local exchange, then that customer would be assigned one of the 4092[41] assignable VIDs.

*Resolution: N:1 forwarding model*

The limitation on the number of S-TAGs is most significant if the customer must be isolated from other customers with his/her own S-TAG; with the N:1 (N ports, 1 VLAN) forwarding model, the low revenue residential customer can be aggregated with far less infrastructural (in terms of S-VIDs used) cost. Therefore, PB that backhauls traffic back to the Internet BNG is not subject to the limitation of 4092 different customers.

*Limitation: time to reconfigure topology in case of link failure (convergence time)*

A serious concern in aggregating using purely Bridged Networks (Ethernet in layers 1 and 2) regards topology reconfigurations. Topology reconfiguration is required whenever a (layer 2) link fails, whether because of port, media or channel failure. IEEE Spanning Tree Protocol (STP) is well known to have unacceptably long convergence times.

*Resolution: simple guarantees on restoration*

There are two principal technology families that support fast topology reconfiguration:

---

[41] The VID field is 12 bits wide and VIDs 0, 1, 2 and FFF are reserved.

1. STP's successors (Rapid Spanning Tree Protocol (RSTP), Multiple Spanning Tree Protocol (MSTP) and Shortest Path Bridging (SPB))[47], and
2. the ITU-T's G.8032 Ethernet ring protection switching (ERPS) [48] and G.8031 (Ethernet linear protection switching) [49].

G.8032 ERPS states simple guarantees on convergence (topology reconfiguration) that are familiar to CSPs: sub-50 ms under specific load conditions and a limit to the number of nodes. There is no similar, evident guarantee with the IEEE standards, although RSTP and MSTP have significantly reduced the time required to adapt active (forwarding) topology to match port state. In the context of the qualitative survey, I have attested to CSPs' perception of longer convergence time with STP and its successors. Academic publications tend to follow suit in this perception of guarantee on availability [50], [51]. It must be added that there appears to be no *empirical* evidence of superiority of ERPS over RSTP, e.g., by comparison of convergence under identical topologies.

Therefore, despite its limitation to ring topology, ITU-T G.8032 provides a topology reconfiguration technique that complements IEEE PB and, furthermore, BBF TR101 Issue 2 VLAN N:1 forwarding model mitigates IEEE PB customer discrimination limit.

*Canonical support for the scenario*

Ethernet aggregation is defined in TR-101 Issue 2[10], wherein the access node is S-VLAN aware and *must* support adding both S-Tag and C-Tag, as necessary. Fig. 32 is a reproduction of [10, Fig. 3]. It serves to support understanding of how PB Q-in-Q can be deployed on the access network: since the access node is S-VLAN aware, and there is no other aggregation technology in use, then upstream traffic is PB Q-in-Q. As stated earlier in this sub-section, S-Tagging of residential customer traffic is likely to follow the N:1 model, where several residential subscribers' ports on the access node are mapped to the same (S-Tagged) VLAN.

*Anecdotal support for the scenario*

A good example of this kind of deployment – and noteworthy corroboration that it is practiced by large-subscriber-base-size CSPs – may be found in [52], wherein it is claimed that "[f]or very large service providers with millions of subscribers this sort of approach normally works well".

*A related technology, that may be on the way out*

A note must be made about an L2 technology that scales better than Provider Bridging, i.e., Provider Backbone Bridging (PBB, IEEE 802.1Q-2022, [47]). However, in two independent e-mail exchanges which I conducted, PBB was described as "all but abandoned which I'm totally OK with. PBB was just basically re-inventing MPLS using Ethernet MAC instead of an MPLS shim" [53], and "PBB-TE … was 'hot' on conferences some years back but the market interest has been limited. Telia has not to my knowledge used the solutions in practice" [54]. The status quo was investigated further through a NANOG posting [55]. The pointed discussion suggests that while some large operators may still use PBB, the technology does not at present have a significant role in expansion of existing aggregation deployments. For this reason, an analyst focusing on representative implementation models has a more significant choice in MPLS-based services than PBB-based ones.

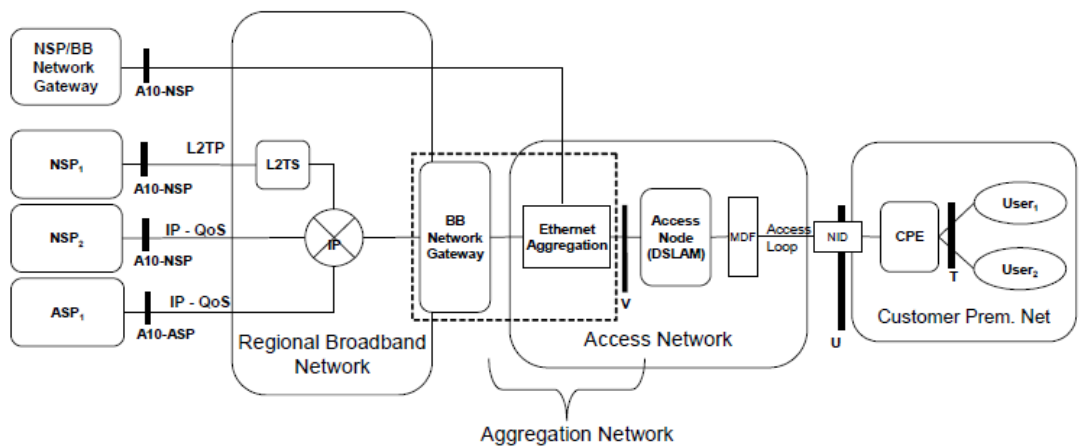

Fig. 32. Simplified schematic of Ethernet-based aggregation of broadband customers [10, Fig. 3]

*A more nuanced reading, leading to other scenarios*

A second look at Fig. 12 and Fig. 13 reveals that, when all the components that aggregate using data-plane MPLS are lumped together, MPLS-based aggregation dominates again. This is less surprising than may appear, as MPLS and PB aggregation are not mutually exclusive. Indeed, the co-existence of MPLS and PB is suggested by the fourth (from top) horizontal bar in Fig. 12 and Fig. 13: "PB Q-in-Q closer to the access with MPLS transport rest of the way back to the service edge". This latter arrangement is one of the "possible combinations of mixing Ethernet and MPLS functions in the various elements, and their corresponding hand-offs" (BBF TR-178 [4, p. 24]). One possible combination will next be presented as a scenario; the combination will then be extrapolated farther to obtain seamless MPLS therefrom.

*A second scenario: PB Q-in-Q closer to the access with MPLS transport rest of the way back to the service edge*

Now, Fig. 32 shows Ethernet aggregation upstream of the access node, but MPLS aggregation could be employed instead. This is the realm of BBF TR-178, and several observations arising out of that standard's provisions are in order.

1. Fig. 7 (taken from TR-178) shows that an external network – network node interface *at layer 2* (E-NNI[42]-L2) may lie anywhere between A10 and V. The means supporting this flexibility is the IP/MPLS-based aggregation (the bottom-most cloud in the diagram). IP/MPLS-based aggregation supports several L2 artefacts, two of which are VPWS and VPLS (see RFC 4664 [57]).

---
[42] E-NNI is used here as defined in MEF 26.2 [56].

2. BBF TR-178 extends TR-101 Issue 2 through consideration of two new access nodes: an MPLS-enabled access node, and a BNG-embedded access node. These access nodes obviate the need for Ethernet aggregation upstream of the V RP.
3. An aggregation implementation that divides the end-to-end path within a CSP's network, into two or more parts, such that:
    i. one (or more) parts employs provider bridging, and
    ii. the other part(s) employs MPLS

    can be referred to as "segmented" MPLS.

As an example of these nuances, a case is now proposed and illustrated in Fig. 33. The context is the following:

1. a single CSP's network's footprint within a single metro area;
2. layer 2 artefacts only are shown;
3. six units of localized real estate, each assigned to a serving area[43];
4. by virtue of its localization, the real estate is a local exchange (LE), or central office, or hub, in the hierarchy of real estate premises;
5. each LE/CO/hub houses:
    a. an Ethernet access node (EAN) for multiplexing of residential subscribers;
    b. zero or more MEF-compliant switches (UNI-Ns) for termination of business subscribers, and
    c. an MPLS PE device;
6. an Internet BNG.

Summarizing, this is a metro area network, and Fig. 33 focuses on the layer 2 architecture of the junction of the edge of the access network, and the edge of the metro-core (which is where the Internet BNG lies). Note that the topology of aggregation, as well as any intermediate real estate, is abstracted for the purpose of clarifying the operation of MPLS.

Each MPLS PE device establishes an MPLS single-segment pseudowire (SS-PW)[44] [24]. Suppose that a business customer attaches to the access node device through an IEEE 802.3 AC, and that the link is a trunk (therefore carrying L2 PDUs with C-VIDs). Furthermore, suppose that the CSP has more than 4092 customers, yet must provide an Ethernet Service Layer. The S-VIDs distinguish the customers at the access node devices, but there are not enough S-VIDs to serve the customer base. The solution here is to use MPLS labels to distinguish the customers between the ingress access node and the egress access node. At the egress access node, the customers can be distinguished using the S-VIDs.

---

[43] There is some redundancy in this phrase, since assignment to a serving area implicitly incurs localization. However, the redundancy is tolerable in so far as it serves to emphasize the geographical scope of the real estate concerned and supports better understanding of the hierarchical arrangement of real estate and network infrastructure.

[44] Note that this is *not yet* operation between two MPLS-enabled access nodes. These access nodes are still Ethernet Access Nodes (EANs).

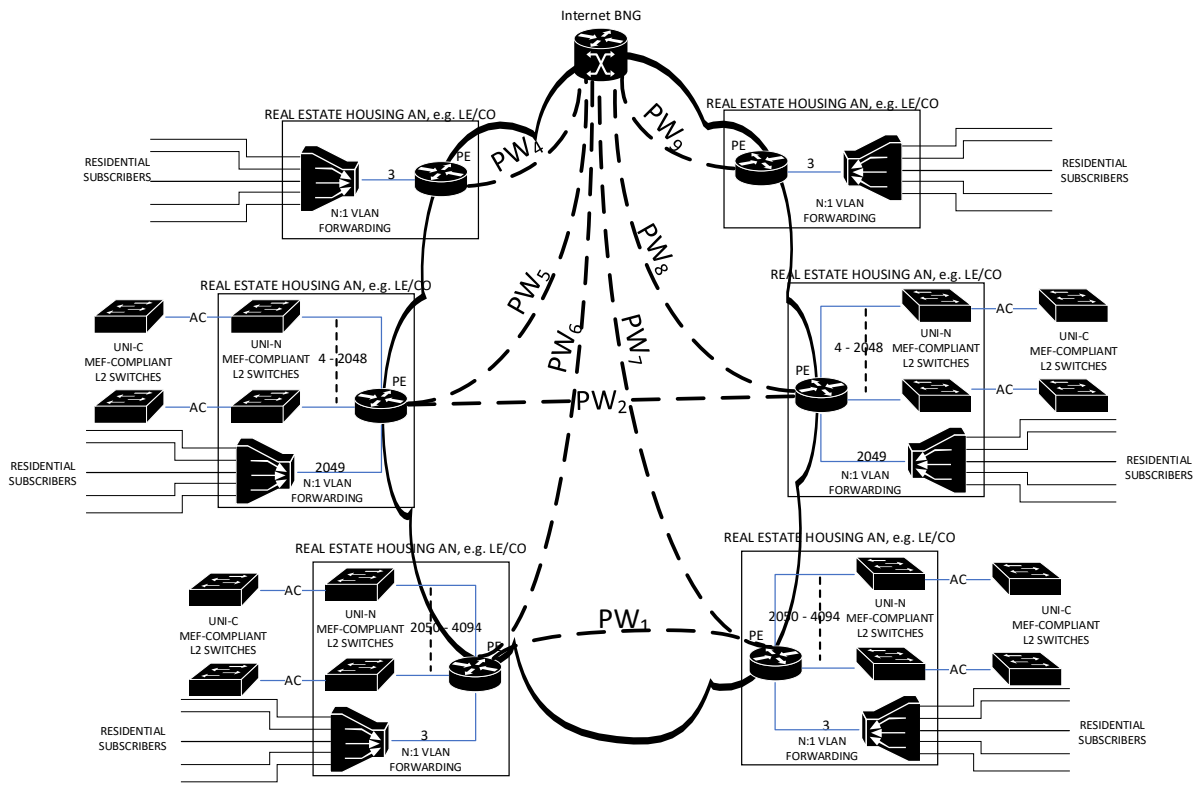

**Fig. 33.** Supporting more than 4092 ETH service layer customers with an MPLS PE in the CSP's localized real estate.

*A third scenario: seamless MPLS*

Fig. 12 and Fig. 13 shows that seamless MPLS dominates aggregation in the smaller subscriber-base CSPs (NOG sample), all the way from the access node, into the metro core. That this dominance emerges within the NOG sample is another indication of smaller-subscriber-base CSPs' agility in the activities of innovation. While seamless MPLS is not radically different in terms of layer 2 architecture, it requires adoption of a newer type of access node: the MPLS-enabled access node. It seems reasonable to expect that the higher capital expenditure and the higher personnel re-training involved would slow the rate at which larger subscriber-base CSPs can adopt seamless MPLS.

Some clarification of the difficulty involved may be obtained from TR-178 (see [4, Fig. 25], reproduced below as Fig. 34), which illustrates the functions embedded in the two types of access node. While the EAN function that interfaces with metro-aggregation at the V RP is the PB functional unit L2F-E [34], the MPLS-enabled AN interfaces at the MPLS adaptation/encapsulation function L2A-M (a function of the LER). The AN is a new device, requiring upgrade, which is better expedited at low unit multiples.

Having established the first scenario as somewhat legacy, yet widely disseminated, and the second scenario as an intermediate step, seamless MPLS has emerged as a desirable next-generation architecture for aggregation. **Error! Reference source not found.** illustrates the implementation. Note that here, the assignment of S-Tags is provided within the MPLS-enabled access node's L2A-E function, *at the service of the Carrier Ethernet Network* (CEN), to support provision of the EVC artefact for customers.

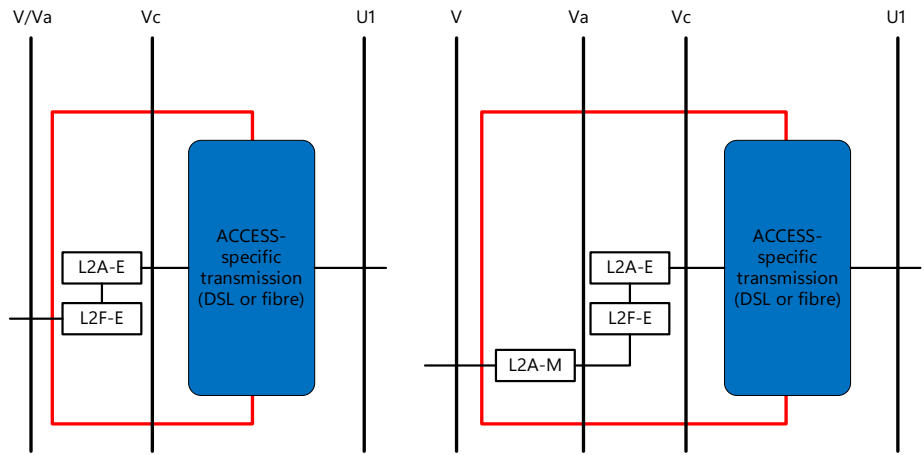

**Fig. 34.** An illustration facilitating a comparison of the functions embedded in the EAN and the MPLS-enabled AN [4, Fig. 25]

It should be noted that the abstraction of metro-aggregation is necessary to focus upon layer 2+ artefacts. Now, it is conceivable, and reasonable, that a single metro area may require no more than a single control domain[45], while remaining within limit of convergence time in the event of link failure. Therefore, the PWs shown in both Fig. 33 and **Error! Reference source not found.** are likely to be single-segment pseudowires (SS-PW), spanning access, metro and core within a single metro area. Notwithstanding the validity of the use case illustrated in **Error! Reference source not found.**, it must be emphasized that the greatest gain in operational simplicity is obtained when an inter-metro PW (or other L2 construct) spanning access-to-access networks, is established with the aid of seamless MPLS.

2) Aggregation at Layers 0 and 1

Both technology and topology are of interest here; with the surveys' results, the most adopted technologies at layers 0 and 1 can be identified[46]. Technologies thus identified will be applied in conjunction with technologies identified earlier in layers 2+ to establish two scenarios.

Provider Bridging, from V to A10

Table VIII shows the stack of technologies in an implementation of an Ethernet access node (EAN) in a ring metro-aggregation arrangement of LEs/COs/hubs. The corresponding schematic, suitably overlaid with reference points, is shown in Fig. 36.

The LE/CO/hub includes other devices: a number of other access nodes (optical line terminals) lie just upstream of the ODN's S/R RP. Moreover, a mobile network operator (MNO, one type of CSP) has co-located[47] its 5G distributed unit (gNB-DU) and centralized unit (gNB-CU) in the real estate.

---

[45] A common and valid example of the scope of a control domain is that within which a set of routers operate in the same IGP (OSPF, IS-IS, say) domain.
[46] The surveys convey some knowledge of topology too, but better confidence in the results requires complementing knowledge thus gained with knowledge gained from case studies.
[47] The MNO may be a division of the same organization, or it may be a different organization that has purchased aggregation from a CSP operating an aggregation network in the metro area.

Both the OLTs and the 5G functional units are uplinked to an Ethernet aggregating device. This is shown separately to accommodate the number of ANs shown in the schematic, but it may well be an integral part of the chassis that includes (the subscriber-facing side of) the access network multiplexer /demultiplexer. Here, this device is the object of attention. Note that Table VIII shows layers that span the range from optical media (layer 0) to layer 2.

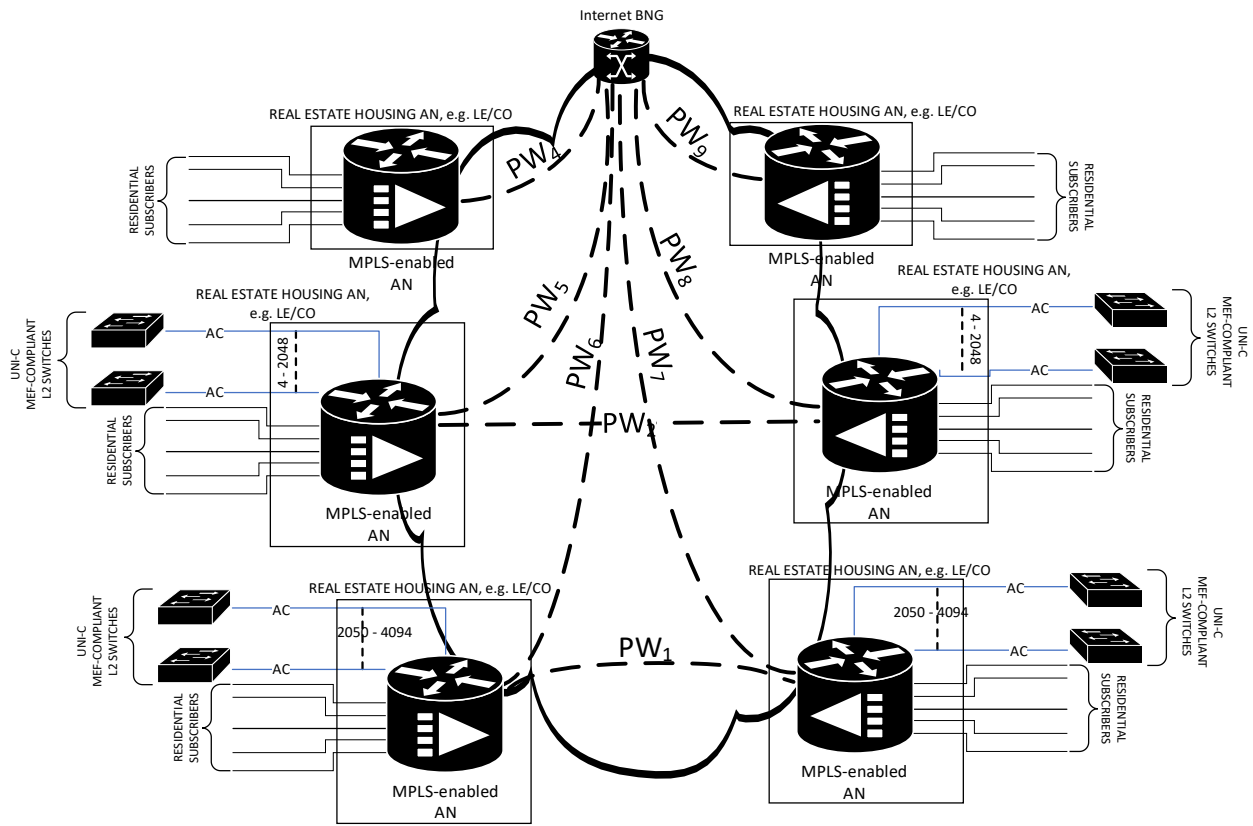

**Fig. 35. An implementation of seamless MPLS**

Consider the MNO's operations. Within the 5GS (Fifth Generation (mobile communications) System), the NG (logical) RP lies upstream of the gNB-CU and downstream of the 5GC; the entire span of the metro-aggregation network is abstracted. *As a consequence of* this abstraction, the protocol stack at the NG-U [58, Fig. 4.3.1.1–1] must be carried unaltered, collapsing the entire metro-aggregation stack into the equivalent of a layer 2 construct (either a link or a MAC Bridge) between the gNB-CU and a UPF in the 5GC (Fifth Generation Core). Here, the carrier's aggregation switch linked to the gNB-CU is the MEF UNI-N device. It does not have MPLS capability, but supports Provider Bridging, with stacked VLANs. and the PDUs on the links involved in switching backhauled traffic, are shown in Fig. 37. Note that although a C-VID is shown, the MNO does not need to deliver a tagged Service Frame; an untagged one is supported in the MEF service set.

MPLS backhaul – a case of recursive transport: Ethernet over MPLS over Ethernet

The concept referred to by the terms underlay and overlay is well-known, but the ITU-T's standards on functional architecture of transport networks[48] use more accurate nomenclature in defining (partition-able) layers which may be vertically traversed in a recursive manner[49], down to the physical media layer [8] (or, specifically for optical networks: the optical media layer [60]). Understanding of Ethernet's role within transport is unsound without an understanding of how it can occupy several layers in a stack of transport technologies.

Table VIII shows the stack of technologies in an implementation of a label edge router (LER) in an access node in a metro-aggregation arrangement. The corresponding schematic, suitably overlaid with reference points, is shown in Fig. 38. As with the previous exposition, the LE/CO/hub includes other devices; the OLTs of various technologies are kept as they were, and again, a mobile network operator (MNO, one type of CSP) has co-located[50] its 5G distributed unit (gNB-DU) and centralized unit (gNB-CU) in the AN's real estate. Once again, the OLTs and the 5G functional units are uplinked to an aggregating device: in this particular case, it is an MPLS-capable device: it is as an LER. Here, the LER is the object of attention; again, layers span the range from optical media (layer 0) to MPLS (layer 2+).

To illustrate this second case better, consider again the MNO's operations. As a consequence of abstraction of all networking entities between the gNB-CU and the UPF, the protocol stack of the UPF at the NG-U [58, Fig. 4.3.1.1–1] must be carried unaltered, collapsing the entire metro-aggregation stack into the equivalent of a layer 2 construct. Here, however, the layer 2 construct is an MPLS artefact (once again: either a link – say a VPWS – or a MAC Bridge – a VPLS) between the gNB-CU and a UPF in the 5GC. The PDUs on the links involved in switching backhauled traffic, are shown in Fig. 39: the MNO is provided with a MEF EVC service with a MEF UNI at the interface between the two physical Ethernet ports linked up within the carrier CSP's[51] access node's real estate.

The specific nature of the MPLS service (itself stacked below an MPLS transport label) is secondary; it could be a VPWS, a VPLS or a BGP EVPN. However, this serves to illustrate the recursive nature of transport: the MNO's Ethernet frame is carried over an MPLS backhaul, which is itself transported over a 10GBASE-ER Ethernet physical layer.

---

[48] Both G.805 [8] – the connection-oriented version – as well as the later G.800 [59] that addresses connectionless as well as connection-oriented transport.

[49] The lowest layer is the transmission media layer network, but this is composed of section layer networks, which include all functions (e.g., transceivers and amplifiers) necessary for information transfer, and the physical media layer network, which is composed of one of the following (a) the fibre, or a channel within the fibre, (b) metallic wire, or a channel therein, and (c) a wireless radio frequency channel.

[50] The MNO may be a division of the same organization, or it may be a different organization that has purchased aggregation from a CSP operating an aggregation network in the metro area.

[51] A CSP is a carrier when it provides a transport service to another CSP, as is the case here.

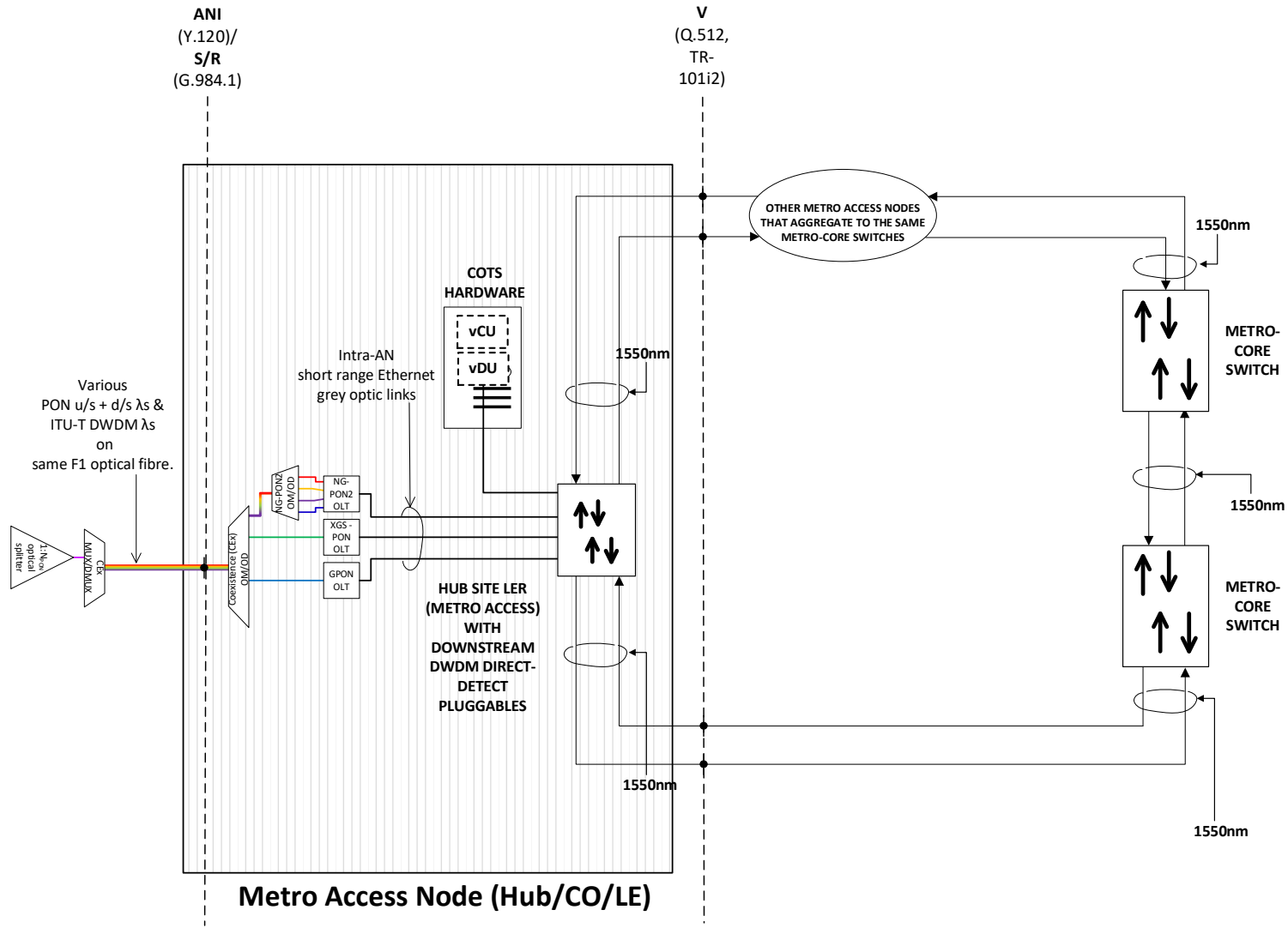

Fig. 36. PB backhaul from access node to metro-core over an Ethernet ring topology

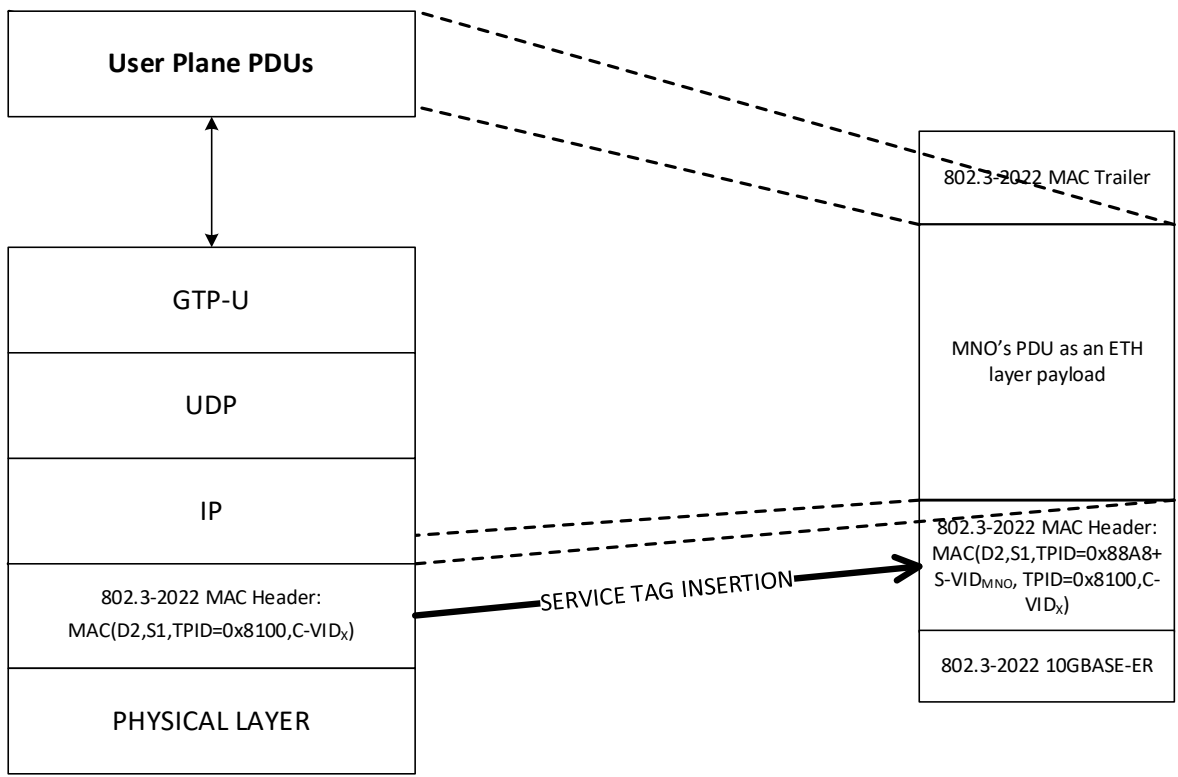

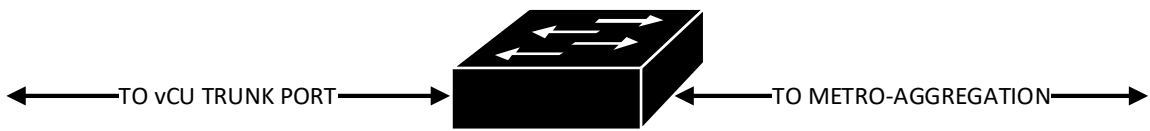

**Fig. 37.** The recursive nature of transport, in the context of 5G backhaul, using IEEE 802.1 Q Provider Bridging

| Layer | Component Type | Descriptor | Comment |
|---|---|---|---|
| | Table VII | PB TECHNOLOGY STACK ON METRO ACCESS NODE AGGREGATION SWITCH IN HUB/CO/LE | |
| **Optical media layer** | Function | Ring of one pair of single-mode physical optical fibre | G.652 |
| **Physical** | Function | IEEE 802.3-2022 10GBASE-ER | Wavelength 1550nm, Link <=40km |
| | Interface | XGMII | PHY-MAC Interface |
| **Link** | Function | IEEE 802.3-2022 Media Access Method Specific Functions | Specific IEEE 802 medium access methods, e.g., for 802.3-2022 |
| | Function | IEEE 802.1 Q-2022 Media Access Method Dependent Convergence Functions | Transforms media-access-method independent functions into media-access-method specific functions |
| | Interface | IEEE 802.1AC-2016 Internal Sublayer Service | Between Media-Access-Method-Dependent-Convergence Functions & Media-Access-Method-Independent Functions |
| | Function | IEEE 802.1Q-2022 Media Access Method Independent Functions | Two interfaces:<br>(a) EISS towards MAC Relay Entity<br>(b) MS towards LLC |
| | Interface | IEEE 802.1Q-2022 Enhanced Internal Sublayer Service | Between Media-Access-Method-Independent Functions & MAC Relay Entity Bridge Port |
| | Function | IEEE 802.1Q-2022 MAC Relay Entity | |
| | Interface | IEEE 802.1Q-2022 MAC Service | Between Media-Access-Method-Independent Functions & LLC |
| | Function | IEEE 802-2014 LLC | Maps higher layer protocols to the MSAP according to EtherType or SNAP addresses |
| | Interface | IEEE 802-2014 LSAP | Between RSTP and LLC |
| | Function | IEEE 802.1Q-2022 RSTP | Avoids loop formation in the ring |

| Layer | Component Type | Descriptor | Comment |
|---|---|---|---|
| Table VIII | | MPLS BACKHAUL TECHNOLOGY STACK ON METRO ACCESS NODE LER (PROVIDER EDGE) IN HUB/CO/LE | |

| Layer | Component Type | Descriptor | Comment |
|---|---|---|---|
| **Optical media layer** | Function | Ring of one pair of single-mode physical optical fibre | G.652 |
| **Physical** | Function | IEEE 802.3-2022 10GBASE-ER | Wavelength 1550nm, Link <=40km |
| | Interface | XGMII | PHY-MAC Interface |
| **Link** | Function | IEEE 802.3-2022 Media Access Method Specific Functions | Specific IEEE 802 medium access methods, e.g., for 802.3-2022 |
| | Function | IEEE 802.1 Q-2022 Media Access Method Dependent Convergence Functions | Transforms media-access-method independent functions into media-access-method specific functions |
| | Interface | IEEE 802.1AC-2016 Internal Sublayer Service | Between Media-Access-Method-Dependent-Convergence Functions & Media-Access-Method-Independent Functions |
| | Function | IEEE 802.1Q-2022 Media Access Method Independent Functions | |
| | Interface | IEEE 802.1Q-2022 MAC Service | Between Media-Access-Method-Independent Functions & LLC |
| | Function | IEEE 802-2014 LLC | Maps higher layer protocols to the MSAP according to EtherType or SNAP addresses |
| | Interface | IEEE 802-2014 LSAP | Between MPLS and LLC |
| **Above Link, below Network** | Function | RFC 3031 MPLS | TR-178-compliant MPLS enabled access node |

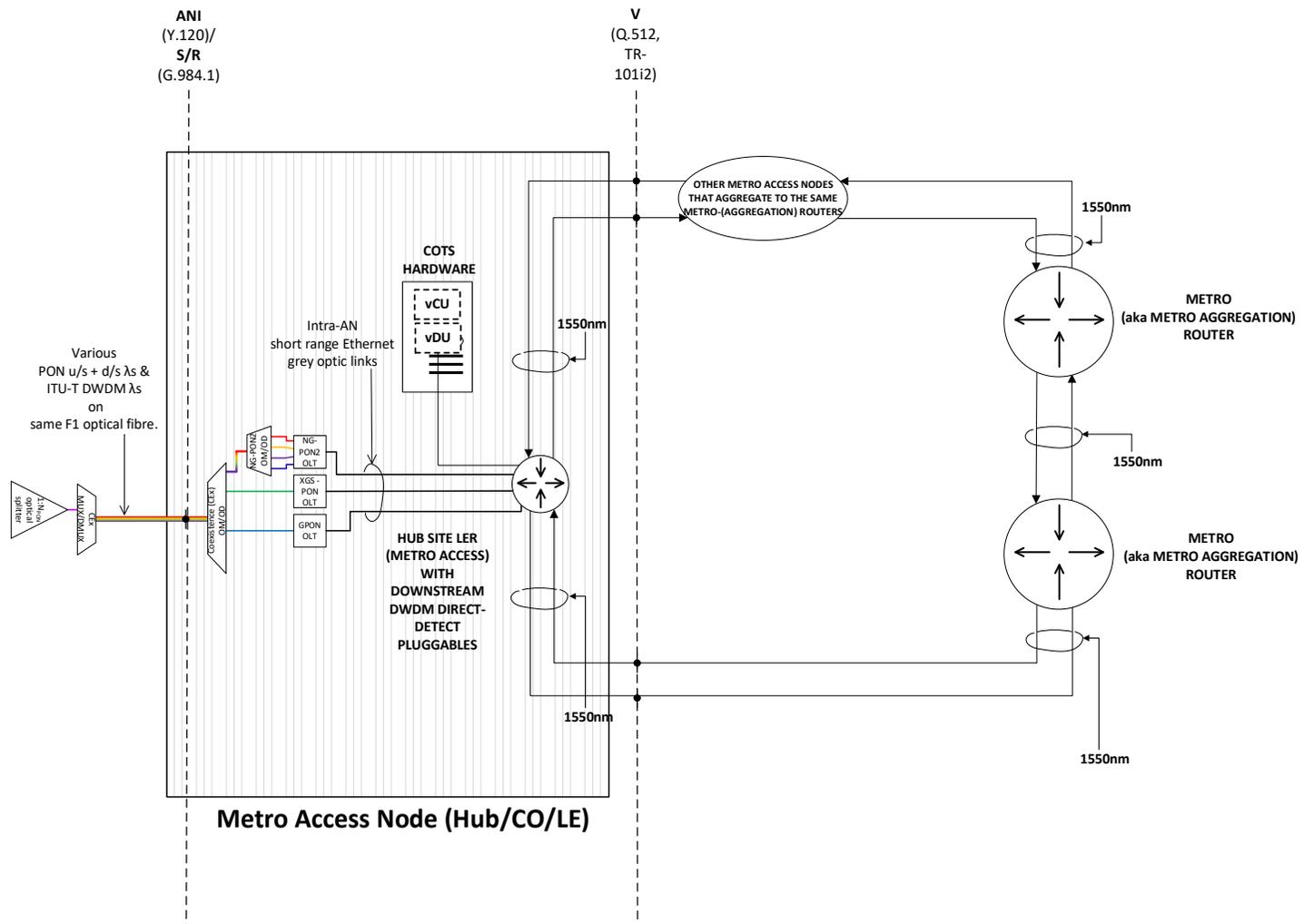

**Fig. 38. MPLS backhaul from access node to metro-core over a ring topology**

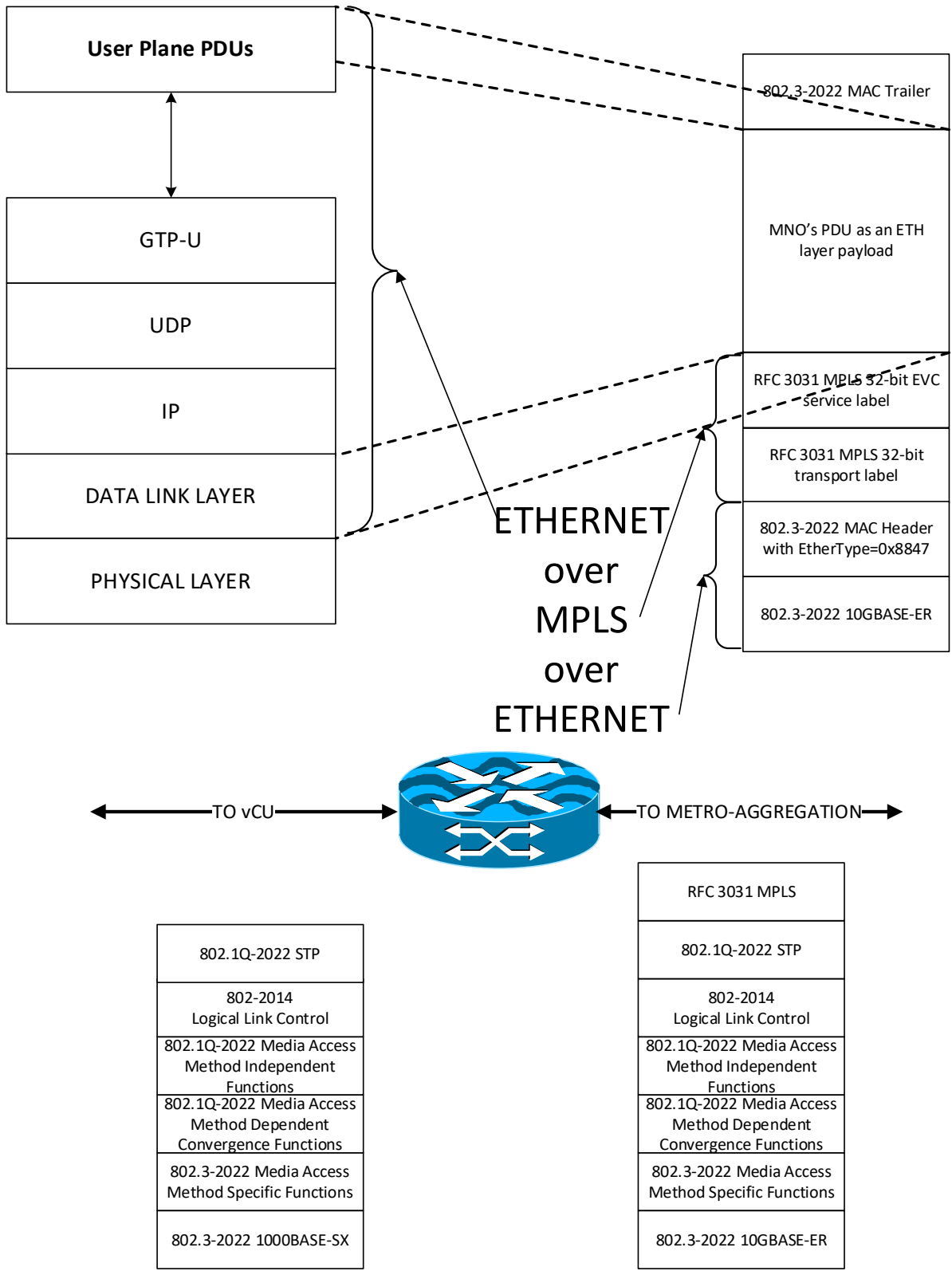

**Fig. 39.** The recursive nature of transport, in the context of 5G backhaul, using MPLS service, e.g., VPLS, VPWS or BGP EVPN

# 7. Conclusion

Our analysis has interpreted the data and elicited several saliencies, e.g.: (1) implementations of metro area networks have been found to bear a relationship to the subscriber base size. This was seen in both choice of access technology as well as aggregation technology. CSPs with larger subscriber base sizes still serve a majority of their subscribers with copper-based access technology, and they are more likely to use both PB and MPLS in aggregation. CSPs with smaller subscriber base sizes serve a majority of their subscribers with ODNs. (2) Selection of layer 0 and layer 1 technologies is guided by the rule of simplicity: if a dark fibre can be spared to implement a medium-haul link between two network elements, then it will be preferred to an optical wavelength in a DWDM + ROADM optical network. OTN presents yet another overlay, dividing the optical wavelength into sub-lambda channels. Operators with smaller subscriber base size are emphatic in favour of avoiding OTN in future aggregation, while operators with larger subscriber base size are less so, but still inclined significantly towards avoiding OTN. (3) Compared with packet-based slicing of transport networks, OTN again emerges as a distant second best in both samples. (4) DAA is another distinguishing aspect; large subscriber-base size CSPs have deployed, or are in the process of deciding upon deployment of remote node. Smaller subscriber-base size CSPs are far less likely to consider deploying remote nodes. Overall, remote OLT and remote MAC-PHY node are the most popular selections. (5) Remote node deployment for multi-access edge computing closely parallels remote node deployment for DAA, indicating that these two aspects of metro area networks are likely to be co-deployed. More generally: beyond these immediate returns, this study posits a point-in-time representation of implementations of CSPs' metro area networks. Two streams of development issue forth from this work. (1) A regular sampling of CSPs can be undertaken, to support understanding of what technologies are succeeding in the field and how implementations are evolving. A regular sampling also supports the robustness of the method by adding further datasets. (2) A richer set of implementational models can be produced out of the dataset, adding more scenarios to those made available here. The ultimate use which the researcher might make of these models seems varied.

Perhaps the ideal destination for all parties, whether CSPs or analysts from academia or industry, would be (a) an SDN platform, (b) informed by implementational models, (c) complemented by YANG [61] NE models that include configuration and notification data that regards increasingly diverse aspects of the NEs' operation and (d) interacting with the data plane using NETCONF [62]. One opportunity awaiting a taker is that of YANG models that account for energy consumption, since much work has been done in ETSI's Green Abstraction Layer (GAL) standards [63], [64]. YANG is concerned with NETCONF data models and protocol operations; these are formally known in RFC 6241 as the Content and Operations layers respectively. As these layers are precisely GAL's scope, GAL can be integrated in the mainstream as a YANG data model for power management. That would go a long way towards meeting the energy analyst's need for a standardized method for data collection.

## LIST OF ACRONYMS

| | |
|---|---|
| 5G | Fifth Generation (mobile communications) |
| 5GC | Fifth Generation Core |
| 5GS | Fifth Generation (mobile communications) System |
| AC | Attachment Circuit |
| ADSL | Asymmetric Digital Subscriber Line |
| AFNOG | AFrica Network Operators' Group |
| AN | Access Node (or Network, according to context) |
| API | Application Programming Interface |
| APOPS | Asia Pacific OPeratorS Forum |
| AS | Autonomous System |
| ATM | Asynchronous Transfer Mode |
| AUSNOG | AUStralian Network Operators' Group |
| BBF | Broadband Forum |
| BGP | Border Gateway Protocol |
| BGP-LS | Border Gateway Protocol - Link State |
| BGP-LU | Border Gateway Protocol - Labeled Unicast |
| BNG | Broadband Network Gateway |
| C-VID | Customer VLAN Identifier |
| CCAP | Converged Cable Access Platform |
| CCIE | Cisco Certified Internetwork Expert |
| CE | Customer Edge |
| CEN | Carrier Ethernet Network |
| CMTS | Cable Modem Termination System |
| CO | Central Office |
| CSP | Communications Service Provider |
| CSR | Cell-Site Router |
| CU | Centralized Unit |
| CWDM | Coarse Wavelength Division Multiplexing |
| DAA | Distributed Access Architecture |
| DCI | DataCentre Interconnect |
| DCSG | Disaggregated Cell-Site Gateway |
| DENOG | German Network Operators' Group |
| DFD | Data-Flow Diagram |
| DOCSIS | Data Over Cable Service Interface Specification |
| DSL | Digital Subscriber Line |
| DSLAM | Digital Subscriber Line Access Multiplexer |
| DU | Distributed Unit |
| DWDM | Dense Wavelength Division Multiplexing |
| EAN | Ethernet Access Node |
| EAS | Energy Aware State |
| ENOG | Eurasia Network Operators' Group |
| EPC | Evolved Packet Core |
| ERPS | Ethernet Ring Protection Switching |
| ESNOG | Spain Network Operators' Group (Grupo de Operadores de Red Españoles) |
| ETSI | European Telecommunications Standards Institute |
| EVC | Ethernet Virtual Connection |
| EVPN | Ethernet Virtual Private Network |
| FPGA | Field Programmable Gate Array |
| FRNOG | France Network Operators' Group |
| FRR | Fast Re-Route |
| FWA | Fixed Wireless Access |
| GAL | Green Abstraction Layer |
| GALv2 | Green Abstraction Layer version 2 |
| GII | Global Information Infrastructure |
| gNB | next generation NodeB |
| GPON | Gigabit Passive Optical Network |
| HCIF | Human-Computer Interfacing Functions |
| HFC | Hybrid Fibre Coaxial |
| HHP | HouseHolds Passed |
| IDNOG | InDonesia Network Operators' Group |
| IEEE | Institute of Electrical and Electronic Engineers |
| IETF | Internet Engineering Task Force |
| IGP | Interior Gateway Protocol |
| INNOG | India Network Operators' Group |
| IS | Information Systems |
| ISG | Industry Specification Group |
| IS-IS | Intermediate System – Intermediate System |
| ISO | International Organization for Standardization |
| ISP | Internet Service Provider |
| IT | Information Technology |
| ITNOG | Italian Network Operators' Group |
| ITU | International Telecommunication Union |
| ITU-T | International Telecommunication Union Telecommunication Standardization Sector |
| JANOG | Japan Network Operators' Group |
| LACNOG | Latin American & Caribbean region Network Operators' Group |
| LAN | Local Area Network |
| LE | Local Exchange |
| LER | Label Edge Router |
| LFA | Loop-Free Alternate |
| LSP | Label-Switched Path |
| MAN | Metro Area Network |
| MEC | Multi-access Edge Computing |
| MEF | Metro Ethernet Forum |
| MPLS | MultiProtocol Label Switching |
| MSBN | Multi-Service Broadband Network |
| MSTP | Multiple Spanning Tree algorithm and Protocol |
| NANOG | North American Network Operators' Group |
| NE | Network Element |
| NETCONF | Network Configuration Protocol |
| NFV | Network Function Virtualization |
| NFVI | Network Function Virtualization Infrastructure |
| NG-PON | Next-Generation Passive Optical Network |
| NMS | Network Management System |
| NOG | Network Operators Group |
| NSP | Network Service Provider |
| OADM | Optical Add-Drop Multiplexer |
| ODN | Optical Distribution Network |
| OLS | Optical Line System |
| OLT | Optical Line Terminal |
| ONT | Optical Network Terminal |
| ONU | Optical Network Unit |
| OPEX | Operational Expenditure |
| OPS | Optical Protection Switching |
| OSFP | Octal Small Form factor Pluggable |
| OSI | Open System Interconnection |
| OSPF | Open Shortest Path First |
| OTN | Optical Transport Network |
| OTT | Over-The-Top |
| OvS | Open virtual Switch |
| PAD | Problem-Approach-Development |
| PB | Provider Bridging |
| PBB | Provider Backbone Bridging |
| PDU | Power Distribution Unit or Protocol Data Unit, according to context |
| PE | Provider Edge |
| PON | Passive Optical Network |
| PSTN | Public Switched Telephone Network |
| PTT | Postal, telegraph, and telephone service |
| QoS | Quality of Service |
| QSFP-DD | Quad Small Form factor Pluggable – Double Density |
| RAM | Random Access Memory |

*List of Acronyms*

| | |
|---|---|
| RAN | Radio Access Network |
| RFC | Request For Comments |
| RG | Residential Gateway |
| ROADM | Reconfigurable Optical Add-Drop Multiplexer |
| RP | Reference Point |
| RPI-N | Reference Point for Interconnection – Network |
| RPN | Remote PHY Node |
| RSTP | Rapid Spanning Tree algorithm and Protocol |
| RU | Radio Unit, or Research Unit, or Rack Unit, depending on context |
| S-VID | Service VLAN Identifier |
| SAFNOG | South African Network Operators' Group |
| SANOG | South Asia Network Operators' Group |
| SDN | Software-Defined Networking |
| SGA | SG Analytics |
| SLA | Service Level Agreement |
| SONET/SDH | Synchronous Optical Network/Synchronous Digital Hierarchy |
| SP | Service Provider |
| SR | Segment Routing |
| STB | Set-Top Box |
| STP | Spanning Tree Protocol |
| SWINOG | SWIss Network Operators' Group |
| TDM | Time Division Multiplexing |
| TE | Traffic Engineering |
| TI-LFA | Topology-Independent Loop-Free Alternate |
| TN | Telecommunications Network |
| UKNOF | United Kingdom Network Operators' Forum |
| UNI | User-Network Interface |
| URL | Uniform Resource Locator |
| URLLC | Ultra-Reliable Low Latency Communication |
| vCMTS | virtual Cable Modem Termination System |
| VLAN | Virtual Local Area Network |
| vOLT | virtual Optical Line Terminal |
| VPLS | Virtual Private LAN Service |
| VPN | Virtual Private Network |
| VPWS | Virtual Private Wire Service |
| WDM | Wavelength Division Multiplexing |
| XG-PON | Ten Gigabit Passive Optical Network |
| XGS-PON | Ten Gigabit Symmetrical Passive Optical Network |
| YANG | Yet Another Next Generation |